\documentclass[reqno]{amsart}

\usepackage{booktabs} 
\usepackage{IEEEtrantools}
\usepackage{amssymb,latexsym,amsfonts,amsmath}
\usepackage{graphicx}
\usepackage{mathrsfs}
\usepackage{dsfont}
\usepackage{paralist}
\usepackage{eso-pic}
\usepackage{epsfig}
\usepackage{mathtools}
\usepackage{epstopdf}
\usepackage{lipsum}
\usepackage{cite}
\usepackage{booktabs}
\usepackage{array}
\usepackage{tikz}
\usetikzlibrary{calc,shapes,arrows}
\usepackage{caption}
\usepackage{subcaption}

\usetikzlibrary{calc,shapes,arrows}

\usepackage{algorithm}
\usepackage{algorithmic}

\usepackage{xspace}

\usepackage{etoolbox}

\newtheorem{theorem}{Theorem}[section]
\newtheorem{lemma}[theorem]{Lemma}

\newtheorem{problem}[theorem]{Problem}

\newtheorem{corollary}[theorem]{Corollary}

\newtheorem{definition}[theorem]{Definition}
\newtheorem{example}[theorem]{Example}

\newtheorem{remark}[theorem]{Remark}
\newtheorem{assumption}{Assumption}
\numberwithin{equation}{section}

\newcommand{\R}{{\mathbb{R}}}

\newcommand{\B}{{\mathbb B}}

\newcommand{\N}{{\mathbb{N}}}

\newcommand{\norm}[1]{\lVert#1\rVert}

\newcommand\blfootnote[1]{%
	\begingroup
	\renewcommand\thefootnote{}\footnote{#1}%
	\addtocounter{footnote}{-1}%
	\endgroup
}

\newcommand{\Let}{:=}

\begin{document}
	
\begin{abstract}
This paper is concerned with a compositional approach for the construction of control barrier certificates for large-scale interconnected stochastic systems while synthesizing hybrid controllers against high-level logic properties. Our proposed methodology involves decomposition of interconnected systems into smaller subsystems and leverages the notion of \emph{control sub-barrier certificates} of subsystems, enabling one to construct control barrier certificates of interconnected systems by employing some $\max$-type small-gain conditions. The main goal is to synthesize hybrid controllers enforcing complex logic properties including the ones represented by the accepting language of deterministic finite automata, while providing probabilistic guarantees on the satisfaction of given specifications in bounded-time horizons. To do so, we propose a systematic approach to first decompose high-level specifications into simple reachability tasks by utilizing automata corresponding to the complement of specifications. We then construct control sub-barrier certificates and synthesize local controllers for those simpler tasks and combine them to obtain a hybrid controller that ensures satisfaction of the complex specification with some lower bound on the probability of satisfaction. To compute control sub-barrier certificates and corresponding local controllers, we provide two systematic approaches based on sum-of-squares (SOS) optimization program and counter-example guided inductive synthesis (CEGIS) framework.  We finally apply our proposed techniques to two physical case studies.
\end{abstract}

\title[Compositional Construction of Barrier Certificates for Large-Scale Stochastic Systems]{From Small-Gain Theory to Compositional Construction of Barrier Certificates for Large-Scale Stochastic Systems}

\author{Mahathi Anand$^{1*}$}
\author{Abolfazl Lavaei$^{2*}$}
\author{Majid Zamani$^{3,1}$}
\blfootnote{\small *Both authors have contributed equally.}
\address{$^1$Department of Computer Science, LMU Munich, Germany}
\email{mahathi.anand@lmu.de}
\address{$^2$Institute for Dynamic Systems and Control, ETH Zurich, Switzerland}
\email{alavaei@ethz.ch}
\address{$^3$Department of Computer Science, University of Colorado Boulder, USA}
\email{majid.zamani@colorado.edu}
\maketitle

\section{Introduction}
Classical control problems can involve checking complex mathematical models against relatively simple properties, such as stability or invariance. On the other hand, the main problem in the formal methods community is to study dynamical models enforcing complex specifications including, but not limited to, safety, reachability and reach-avoid. In particular, in the past few years, formal verification and synthesis of complex stochastic systems against a wide variety of high-level specifications have gained considerable attentions~\cite{Tabuada.2009}.  Many safety-critical scenarios such as power networks, air traffic control, and so on, can be modeled by stochastic control systems, and these types of problems are especially challenging when dealing with large-scale systems with continuous state and input sets.

Existing results on the verification and controller synthesis of large-scale stochastic systems have been widely focused on abstraction-based techniques. Such approaches include probabilistic reachability guarantees for discrete-time stochastic hybrid systems via abstraction~\cite{Abate2008}, game-based abstractions for controller synthesis~\cite{hahn_game-based_2011} in stochastic hybrid automata, and an abstraction-based framework for the synthesis of bounded Markov decision processes against probabilistic computation tree logic (PCTL)~\cite{Lahijanian.2015}. However, these techniques rely on state-space discretization, and accordingly, computational complexity increases exponentially with the dimension of the state space. This issue has been partially alleviated by using
sequential gridding procedures~\cite{EsmaeilZadehSoudjani.2013} and input-set abstraction for incrementally stable stochastic control systems~\cite{Zamani.2017}. 
As an alternative solution proposed in recent years, one can consider a large-scale system as an interconnection of smaller subsystems and employ compositionality techniques for constructing finite abstractions of interconnected systems based on abstractions of subsystems~\cite{lavaei2018ADHS,lavaei2022LSS_J,Lavaei_Survey}. 
More recently, discretization-free approaches via control barrier certificates have been proposed for the verification and synthesis of stochastic systems. Existing results include safety verification of continuous-time stochastic hybrid systems in infinite-time horizons~\cite{prajna2007,Wisniewski.2018}. 
Such verification in infinite time requires a supermartingale condition that implicitly assumes the system's stability at its equilibrium point which is restrictive. In~\cite{steinhardt2012finite}, this condition is generalized to the safety verification of stochastic systems in finite-time domains. Controller synthesis for the finite-time safety of stochastic systems using control barrier certificates is discussed in~\cite{santoyo2019barrier}.
Systematic verification and synthesis techniques against temporal logic specifications for nonlinear systems are provided in~\cite{wongpiromsarn_automata_2016,larscbf} and for Markov decision processes in \cite{ahmadipomdp}. Finite-time  controller synthesis for discrete-time stochastic control systems using barrier certificates against automata representation of temporal logic properties is proposed in \cite{jagtap_formal_2019}. 

The proposed techniques in the aforementioned literature involve restricting control barrier certificates to a certain parametric form, such as exponential or polynomial, by searching for their corresponding coefficients under certain assumptions. Although lower-dimensional systems usually admit such simple control barrier certificates and the corresponding search is relatively easy using existing tools, it may be very difficult (if not impossible) in the case of large-scale systems, and therefore, such techniques become computationally intractable. 

In order to overcome the aforementioned challenge, we propose a compositional framework for the construction of control barrier certificates for large-scale interconnected stochastic systems. The proposed approach involves decomposing a large-scale stochastic system into a number of smaller subsystems of lower dimensions, and searching for control sub-barrier certificates for those subsystems together with corresponding local controllers. By leveraging some $\max$ small-gain conditions, a control barrier certificate and its corresponding controller for the interconnected system can be constructed from control sub-barrier certificates and corresponding local controllers of subsystems. The control barrier certificate is then utilized to establish upper bounds on the probability that interconnected systems reach unsafe regions within finite-time horizons, thereby allowing finite-time verification and synthesis of safety properties.

For synthesizing controllers for more general specifications, we provide a systematic method to decompose a complex property  that can be expressed by an accepting language of a deterministic finite automaton (DFA) into simpler tasks based on the complement automaton of the original specification. Control sub-barrier certificates are then computed for each task along with the corresponding probabilities which can eventually be combined to obtain a lower-bound probability using which the system would satisfy the original specification over a finite-time horizon. Correspondingly, a hybrid controller is achieved for the large-scale interconnected stochastic system that ensures the satisfaction of the given specification. 
We finally apply our proposed results to a \emph{fully-interconnected} Kuramoto network with $100$ \emph{nonlinear} oscillators, and synthesize hybrid controllers to ensure satisfaction of a complex specification given by a deterministic finite automaton.       

Compositional construction of control barrier certificates via small-gain theorem is presented in~\cite{lyu2020smallgain} but in the context of input-to-state \emph{safety} properties for \emph{non-stochastic} interconnected systems with only two subsystems. In comparison, our proposed results are for \emph{stochastic} large-scale systems without putting any restrictions on the number of subsystems. Moreover, we study here a larger class of logic specifications described by deterministic finite automata (DFA). Compositional construction of control barrier certificates for non-stochastic control systems is also presented in~\cite{jagtap_compositional_2020} for enforcing specifications that can be described by deterministic B\"uchi automata (DBA) over \emph{infinite-time horizons}. In comparison, we deal with stochastic control systems and provide \emph{finite-time} horizon guarantees for specifications expressed by DFA.
Compositional construction of control barrier certificates for large-scale stochastic systems is recently discussed in~\cite{mahathi_IFAC2020}. Our current work generalizes~\cite{mahathi_IFAC2020} in two main directions. First and mainly, we do not restrict ourselves to verification and synthesis over simple \emph{safety} specifications and consider a larger class of specifications that can be admitted by accepting languages of DFA. 
As our second contribution, this paper includes a comprehensive fully-interconnected nonlinear case study against complex logic properties expressed by DFA that illustrates the proposed results. 
Compositional construction of control barrier functions for large-scale stochastic systems are presented in~\cite{AmyAutomatica2020_J} but for \emph{continuous-time} stochastic systems with a different compositional technique based on \emph{sum-type} small-gain conditions. Unfortunately, those conditions are conservative as they are all formulated in terms of ``almost" linear gains, which means that subsystems should have a (nearly) linear behavior. Compositional construction of safety controllers for networks of continuous-space POMDPs using control barrier certificates is recently proposed in~\cite{Niloofar_TNCS_2022}.

\section{Discrete-Time Stochastic Control Systems} \label{prelim}

\subsection{Preliminaries}

In this work, we consider the probability space $(\Omega, \mathcal{F}_\Omega, \mathbb{P}_\Omega)$, where $\Omega$ is the sample space, $\mathcal{F}_\Omega$ is a sigma-algebra on $\Omega$ consisting subsets of $\Omega$ as events, and $\mathbb{P}_\Omega$ is the probability measure that assigns probabilities to those events. Random variables introduced in this paper are assumed to be measurable functions of the form $X:(\Omega,\mathcal{F}_\Omega) \rightarrow (S_X,\mathcal{F}_X) $ such that any random variable $X$ induces a probability measure on its space $(S_X,\mathcal{F}_X)$
as $Prob\{A\} = \mathbb{P}_\Omega\{X^{-1}(A)\}$ for any $A \in \mathcal{F}_X.$ The probability measure on $(S_X,\mathcal{F}_X)$ is presented directly without any explicit mention of the underlying probability space or the function $X$. The topological space $S$ is a Borel space if it is homeomorphic to a Borel subset of a Polish space, \textit{i.e.}, a separable and completely metrizable space. 
$\mathcal{B}(S)$ denotes the Borel sigma-algebra which is generated from a Borel space $S$. The map $f: S \rightarrow Y$ is measurable whenever it is Borel measurable.

\subsection{Notations}

We denote the set of real, positive and non-negative real numbers by $\mathbb{R},\mathbb{R}_{>0}$, and $\mathbb{R}_{\geq 0}$, respectively. In addition, $\mathbb{R}^n$ denotes a real space of dimension $n$. We use $\mathbb{N} := \{0,1,\ldots\}$ to represent the set of non-negative integers and $\mathbb{N}_{\geq 1}=\{1,2,\ldots\}$ to denote the set of positive integers. Given $N$ vectors $x_i \in \mathbb{R}^{n_i}$, the corresponding column vector of dimension $\sum_i n_i$ is denoted by $x=[x_1;\ldots;x_N]$. For a vector $x\in\mathbb{R}^{n}$, an infinity norm of $x$ is denoted by $\Vert x\Vert$. Symbols $\mathds{I}_n$, $\mathbf{0}_n$, and $\mathds{1}_n$ denote the identity matrix in $\mathbb R^{n\times{n}}$ and the column vectors in $\mathbb R^{n\times{1}}$ with all elements equal to zero and one, respectively. The identity
function and composition of functions are denoted by $\mathcal{I}_d$ and symbol $\circ$, respectively. Given functions $f_i:X_i\rightarrow Y_i$, for any $i\in\{1,\ldots,N\}$, their Cartesian product $\prod_{i=1}^{N}f_i:\prod_{i=1}^{N}X_i\rightarrow\prod_{i=1}^{N}Y_i$ is defined as $(\prod_{i=1}^{N}f_i)(x_1,\ldots,x_N)=[f_1(x_1);\ldots;f_N(x_N)]$.  For a set $S$, $|S|$ denotes its cardinality and empty set is denoted by $\emptyset$. Given a set $S$ and $P \subset S$, the complement of $P$ with respect to $S$ is given by $S\backslash P= \{x | x \in S, x \notin P\}$. The power set of $S$ is the set of all subsets of $S$ and is denoted by $2^{S}$. We denote the disjunction ($\vee$) and conjunction ($\wedge$) of a Boolean function $f: S \rightarrow \{0,1\}$ over a (possibly infinite) index set $S$ by $\bigvee_{s \in S} f(s)$ and $\bigwedge_{s \in S}f(s)$, respectively. A function $\varphi: \mathbb{R}_{\geq 0} \rightarrow \mathbb{R}_{\geq 0}$ is said to be a class $\mathcal{K}$ function if it is continuous, strictly increasing, and $\varphi(0)=0$. A class $\mathcal{K}$ function $\varphi(\cdot)$ belongs to the class $\mathcal{K}_\infty$ if $\varphi(s) \rightarrow \infty$ as $s \rightarrow \infty$. 

\subsection{Discrete-Time Stochastic Control Systems}

In this paper, we focus on discrete-time stochastic control systems (dt-SCS), as formalized in the following definition.
\begin{definition} \label{def:dtSCS_wo_internal}		
	A \emph{discrete-time stochastic control system} (dt-SCS) is a tuple 
	\begin{equation}\label{eq:dt-SCS}
	\mathfrak{S}=(X,U,\varsigma,f),
	\end{equation}
	where,
	\begin{itemize}
		\item
		$X\subseteq \mathbb R^n$ is a Borel set as the state set of the
		system;
		\item $U\subseteq \mathbb R^m$ is a Borel set as the input set of the system; 
		\item
		$\varsigma$ is a sequence of independent and identically distributed
		(i.i.d.) random variables from a sample space $\Omega$ to the
		measurable space
		$(\mathcal V_\varsigma, \mathcal F_\varsigma)$, namely	 
		$\varsigma:=\{\varsigma(k):\Omega\rightarrow \mathcal V_{\varsigma},\,\,k\in\N\}$;
		\item
		$f:X{\times} U{\times} \mathcal V_{\varsigma} \rightarrow X$ is a measurable function that
		characterizes the state evolution of $\mathfrak{S}$.
	\end{itemize}
\end{definition}

The evolution of the state of dt-SCS $\mathfrak{S}$ for a given initial state $x(0) \in X$, and input sequence $\{\nu(k):\Omega\rightarrow U,\,\,k\in\mathbb N\}$ is described by: 

\begin{equation} \label{Eq_2a}
\mathfrak{S}:x(k+1)=f(x(k),\nu(k),\varsigma(k)).
\vspace{-0.5em}
\end{equation}

A set $\mathcal{U}$ is associated with $U$ as a collection of sequences $\{\nu(k):\Omega\rightarrow U,\,\,k\in\mathbb N\}$, where $\nu(k)$ is independent of $\varsigma(z)$ for any $k,z\in\mathbb N$ and $z\ge k$. For any initial state $a \in X$ and $\nu(\cdot) \in \mathcal{U}$, $x^{a\nu}: \Omega \times \mathbb{N} \rightarrow X$ denotes the solution process of $\mathfrak{S}$ under the input sequence $\nu$ and an initial state $a$. 
We now present history-dependent policies to control dt-SCS in~\eqref{eq:dt-SCS}. 
\begin{definition} \label{controlpolicy}
	For a dt-SCS $\mathfrak S$ as in~\eqref{eq:dt-SCS}, a history dependent policy 
	$\varpi = (\varpi_0,\varpi_1,\ldots)$ is a sequence with functions $\varpi_i : \mathcal{G}_i \rightarrow U$, where $\mathcal{G}_i$ is the set of all $i$-histories $g_i$ that can be defined as $g_i := (x(0),\nu(0),x(1),$ $\nu(1),\ldots,x(i-1),\nu(i-1),x(i))$. Stationary policies are a subclass of history-dependent policies where $\varpi=(\nu,\nu,\ldots)$, $ \nu: X \rightarrow U$. Here, the mapping at any time $i$ only depends on the current state $x(i)$ and is not time-variant.
\end{definition} 

This article is mainly concerned with the controller synthesis for large-scale \emph{interconnected} dt-SCS as in~\eqref{eq:dt-SCS}, that can be considered as compositions of several smaller subsystems. These subsystems consist of \emph{internal} and \emph{external} inputs, as well as outputs, as defined below.
\begin{definition} \label{def:dt-SCS-with-int}
	A dt-SCS with internal inputs is a tuple $\mathfrak{S}=(X,U,W,\varsigma,f,Y,h)$, where $X, U$ and $W$ are Borel sets as the state set, \emph{external} input set and \emph{internal} input set of the system, respectively. $\varsigma$ is a sequence of i.i.d. random variables from a sample space $\Omega$, $f: X \times U \times W \times \mathcal{V}_{\varsigma} \rightarrow X$ is a measurable function characterizing the state evolution of the system, $Y$ is a Borel set as the output set of the system, and $h: X \rightarrow Y$ is a measurable output function that maps states of the system to their outputs $y=h(x)$. 
\end{definition}
Consequently, the dynamics in~\eqref{Eq_2a} is extended accordingly to dt-SCS with internal inputs and outputs and is described by 
\begin{equation}\label{Eq_1a}
\mathfrak S:\left\{\begin{array}{l}x(k+1)=f(x(k),\nu(k),w(k),\varsigma(k)),\\
y(k)=h(x(k)),\\
\end{array}\right. k\in\mathbb N.
\end{equation}
Moreover, we associate with $W$ a set $\mathcal{W}$ to be a collection of sequences $\{w(k):\Omega\rightarrow W,\,\,k\in\mathbb N\}$, where $w(k)$ is independent of $\varsigma(z)$ for any $k,z\in\mathbb N$ and $z\ge k$. Now, for any initial state $ a \in X$, $\nu(\cdot) \in \mathcal{U}$ and  $w(\cdot) \in \mathcal{W}$, the solution process of $\mathfrak{S}$ is denoted by random sequences $x^{a\nu w}:\Omega \times\mathbb N \rightarrow X$ under an internal input $\nu$, an external input $w$, and an initial state $a$. 

\begin{remark}
	Note that the main role of outputs in~\eqref{Eq_1a} is for the sake of interconnection which we will see later. More precisely, we assume that the output map of the interconnected system is identity (\emph{i.e.,} the full-state information is available), as appears in~\eqref{Eq_2a}, mainly for the sake of controller synthesis.
\end{remark}

\section{Control (Sub-)barrier Certificates} \label{barrier}

We first define control barrier certificates for interconnected discrete-time stochastic control systems, borrowed from~\cite{mahathi_IFAC2020}, which will later be used to obtain probabilistic guarantees on the satisfaction of specifications over interconnected systems. 

\begin{definition} \label{cbc}
	Consider an interconnected dt-SCS $\mathfrak{S}= (X,U,\varsigma,f)$ without internal inputs. A function $\mathds B:X \rightarrow \mathbb{R}_{\geq 0}$ is called a control barrier certificate (CBC) for $\mathfrak{S}$ if 
	\begin{align}\label{sys2}
	&\mathds B(x) \leq \eta,\quad\quad\quad\quad\quad\!\forall x \in X_{0},\\\label{sys3}
	&\mathds B(x) \geq \beta, \quad\quad\quad\quad\quad\!\forall x \in X_{u},
	\end{align}  
	and $\forall x\in X$, $\exists u\in U$, such that 
	\begin{align}
	\mathbb{E}&\Big[\mathds B(x(k+1)) \,\,\big|\,\, x(k)=x, \nu(k)=u\Big] \leq \max\Big\{\kappa(\mathds B(x(k))), c\Big\}, \label{cbceq}
	\end{align}
	for a function
	$\kappa\in\mathcal{K}_\infty$, with $\kappa <\mathcal{I}_d$, and constants $\eta,c\in\R_{\geq 0}$ and $\beta\in\R_{> 0}$, with $\beta > \eta$.
\end{definition}

A similar definition, borrowed from~\cite{mahathi_IFAC2020}, is applied for dt-SCS with both external and internal inputs. 

\begin{definition} \label{csbc}
	Consider a dt-SCS with both internal and external inputs as  $\mathfrak{S}=(X,U,W,\varsigma,f,Y,h)$, with sets $X_0, X_u \subseteq X$ as initial and unsafe sets of the system, respectively. A function $\mathds B:X \rightarrow \mathbb{R}_{\geq 0}$ is said to be a control
	sub-barrier certificate (CSBC) for $\mathfrak{S}$ if there exist functions
	$\alpha,\kappa\in\mathcal{K}_\infty$, with $\kappa <\mathcal{I}_d$, $\rho\in\mathcal{K}_\infty\cup\{0\}$, and constants $\eta,c\in\R_{\geq 0}$ and $\beta\in\R_{> 0}$, such that
	\begin{align}\label{subsys1}
	&\mathds B(x) \geq \alpha(\Vert h(x)\Vert^2),\quad\quad\!\forall x \in X,\\\label{subsys2}
	&\mathds B(x) \leq \eta,\quad\quad\quad\quad\quad\quad\forall x \in X_{0},\\\label{subsys3}
	&\mathds B(x) \geq \beta, \quad\quad\quad\quad\quad\quad\forall x \in X_{u}, 
	\end{align}  
	and $\forall x\in X$, $\exists u\in U$, such that $\forall w\in W$,
	\begin{align}
	\mathbb{E}&\Big[\mathds B(x(k+1)) \,\,\big|\,\, x(k)=x, \nu(k)=u, w(k)=w\Big] \leq \max\Big\{\kappa(\mathds B(x(k))), \rho(\|w\|^2),c\Big\}. \label{csbceq}
	\end{align}
\end{definition}

\begin{remark}
	We require condition $\beta > \eta$ in Definition~\ref{cbc} for interconnected systems in order to propose meaningful probabilistic bounds on the satisfaction of specifications using Theorem~\ref{Kushner}. However, we do not ask such a condition in Definition~\ref{csbc} for dt-SCS with internal inputs, since a CSBC does not explicitly provide any probabilistic safety guarantees. In fact, CSBCs as in Definition~\ref{csbc} are only utilized to compute CBCs for the interconnected system, which then provide safety guarantees over the interconnected system (cf. Section~\ref{compositional}).
\end{remark}
Now, employing Definition \ref{cbc}, we provide a theorem, borrowed from~\cite{mahathi_IFAC2020}, that quantifies an upper bound on the probability that an interconnected dt-SCS reaches an unsafe region in a finite-time horizon.

\begin{theorem} \label{Kushner}
	Let $\mathfrak{S}= (X,U,\varsigma,f)$ be an interconnected dt-SCS. Suppose $\mathds B$ is a CBC for $\mathfrak{S}$ and there exists a constant $0<\hat\kappa< 1$ such that function $\kappa \in \mathcal{K}_\infty$ in~\eqref{cbceq} satisfies $\kappa(s)\leq\hat\kappa s$, $\forall s\in\mathbb R_{\geq0}$. Then the probability that the solution process of $\mathfrak{S}$ starts from any initial state $a\in X_0$ and reaches an unsafe region $X_u$ under the controller $\nu(\cdot)$ 
	within finite time steps $k\in [0,T_d)$ is lower bounded as 
	\begin{equation} \label{eqlemma2}
	\mathbb{P}^{a}_{\nu} \Big\{\sup_{0 \leq k < T_d} \mathds B(x(k)) \geq \beta \,\, \big|\,\, a\Big\} \leq \varkappa,
	\end{equation}
	where, 
	\begin{equation*}
	\varkappa=  \begin{cases} 
	1-(1-\frac{\eta}{\beta})(1-\frac{c}{\beta})^{T_d}, &  \quad\text{if } \beta \geq \frac{c} {1-\hat{\kappa}}, \\
	\frac{\eta}{\beta}\hat{\kappa}^{T_d}+\frac{c}{	(1-\hat{\kappa})\beta}(1-\hat{\kappa}^{T_d}), & \quad\text{if } \beta < \frac{c} {1-\hat{\kappa}}.  \\
	\end{cases}
	\end{equation*}	
\end{theorem}

The proof of Theorem \ref{Kushner} is provided in Appendix. The results of Theorem \ref{Kushner} provide upper bounds on the probability that interconnected systems reach unsafe regions in \emph{finite-time} horizons. The proposed results can be extended to \emph{infinite-time} horizons when the constant $c = 0$. This is provided in the following corollary.

\begin{corollary}\label{Kushner1}
	Let $\mathfrak{S}= (X,U,\varsigma,f)$ be an interconnected dt-SCS without internal inputs. Suppose $\mathds B$ is a CBC for $\mathfrak{S}$ such that the constant $c = 0$ in~\eqref{cbceq}. Then the probability that the solution process of $\mathfrak{S}$ starts from any initial state $a\in X_0$ and reaches $X_u$ under the controller $\nu(\cdot)$ (associated with the CBC $\mathds B$ and satisfying condition \eqref{cbceq})  within the infinite time step $k\in [0,\infty)$ is
	\begin{equation*}
	\mathbb{P}^{a}_{\nu} \Big\{\sup_{0 \leq k < \infty} \mathds B(x(k)) \geq \beta\,\,\big|\,\,  a\Big\} \leq \frac{\eta}{\beta}.
	\end{equation*}
\end{corollary}
The proof is similar to that of Theorem~\ref{Kushner} by applying \cite[Theorem 12, Chapter II]{1967stochastic} and is omitted here.

\begin{remark} 
	Note that CBC $\mathds B$ satisfying the condition~\eqref{cbceq} with $c = 0$ is a non-negative supermartingale~\cite[Chapter I]{1967stochastic}.  Although the supermartingale property on $\mathds B$ allows one to provide probabilistic guarantees for \emph{infinite-time} horizons via Corollary~\ref{Kushner1}, it is restrictive in the sense that a supermartingale CBC $\mathds B$ may not exist in general \cite{steinhardt2012finite}. We therefore employ a more general $c$-martingale type condition at the cost of providing probabilistic guarantees for \emph{finite-time} horizons.
\end{remark}	

In the next section, we describe interconnected stochastic control systems as a composition of several stochastic subsystems, and provide compositional conditions under which a CBC of an interconnected system can be constructed from CSBCs of subsystems.

\section{Compositional Construction of CBC}\label{interconnected}

\subsection{Interconnected Stochastic Control Systems}

Suppose we are given $N$ control subsystems 
\begin{align}\label{subsystems}
\mathfrak{S}_i = (X_i,U_i,W_i,\varsigma_i,f_i,Y_i,h_i),\quad i\in \{1,\dots,N\},\,
\end{align} where $X_i \in \mathbb{R}^{n_i}$, $U_i \in \mathbb{R}^{m_i}$, $W_i \in \mathbb{R}^{p_i}$, and $Y_i \in \mathbb{R}^{q_i}$, whose internal inputs and outputs are partitioned as 
\begin{align}\notag
w_i&=[{w_{i1};\ldots;w_{i(i-1)};w_{i(i+1)};\ldots;w_{iN}}],\\\label{config1}
y_i&=[{y_{i1};\ldots;y_{iN}}],
\end{align}
and their output spaces and functions are of the form 
\begin{equation}
\label{config2}
Y_i=\prod_{j=1}^{N}Y_{ij},\quad h_i(x_i)=[{h_{i1}(x_i);\ldots;h_{iN}(x_i)}].
\end{equation}

We call outputs $y_{ii}=x_i$ as \emph{external} ones, whereas outputs $y_{ij}$ with $i\neq j$ are \emph{internal} ones which are used to interconnect stochastic control subsystems. If there exists a connection from $\mathfrak{S}_{j}$ to $\mathfrak{S}_i$, then $w_{ij} = y_{ji}$. Otherwise, the connecting output is considered identically zero, \emph{i.e.,} $h_{ji}\equiv 0$.

\begin{remark}
	The term ``internal" is utilized to refer to those inputs and outputs of subsystems that affect the behavior of other subsystems, \textit{i.e.}, an internal input of a subsystem is affected by an internal output of another one. The term ``external" is employed to describe those inputs and outputs that are not used for constructing the interconnection.
	In this paper, we assume that one has full-state information in order to synthesize controllers, \textit{i.e.,} $h_{ii}(x_i)=x_i$. In the absence of full-state information, the controller synthesis becomes more challenging since one requires the existence of an estimator with some given accuracy. See~\cite{nil_pomdp} for a detailed discussion. Under this assumption, we are able to formulate CSBCs and controllers directly over the actual states of the system.
\end{remark}
We now provide a formal definition of interconnected discrete-time stochastic control systems. 

\begin{definition}
	Consider $N\in\mathbb N_{\geq1}$ stochastic control subsystems $\mathfrak{S}_i=(X_i,U_i,W_i,\varsigma_i,f_i,Y_i,h_i)$, $i\in \{1,\dots,N\}$, with the input-output partition as in \eqref{config1} and \eqref{config2}. The interconnected discrete-time stochastic control system $\mathfrak{S}=(X,U,\varsigma,f)$ is composed of  $\mathfrak{S}_i$, $\forall i\in \{1,\ldots,N\}$, denoted by $\mathcal{I}(\mathfrak{S}_1,\ldots,\mathfrak{S}_N)$ such that $X:=\prod_{i=1}^{N}X_i$, $U:=\prod_{i=1}^{N}U_i$, $\varsigma:=[\varsigma_1;\ldots;\varsigma_N]$, and $f:=\prod_{i=1}^{N}f_{i}$, subjected to: 
	$$\forall i,j\in \{1,\dots,N\},i\neq j\!: ~~~ w_{ji} = y_{ij}, ~~~ Y_{ij}\subseteq W_{ji}.$$
\end{definition}

\subsection{Compositional Construction of CBC for Interconnected Systems}\label{compositional}

In this subsection, we provide a compositional framework for the construction of CBC for $\mathfrak{S}$ using CSBC of $\mathfrak{S}_i$. For each control subsystem $\mathfrak{S}_i, i\in \{1,\dots,N\}$ in~\eqref{subsystems}, suppose there exists CSBC $\mathds B_i$ as defined in Definition \ref{csbc} with functions
$\alpha_i,\kappa_i\in\mathcal{K}_\infty$, with $\kappa_i <\mathcal{I}_d$, $\rho_i\in\mathcal{K}_\infty\cup\{0\}$, and constants $\eta_i,c_i\in\R_{\geq 0}$ and $\beta_i\in\R_{> 0}$. Now we present the following small-gain assumption that is essential for the compositional construction of CBC for $\mathfrak{S}$.

\begin{assumption}\label{Assump: Gamma}
	Assume that $\mathcal{K}_\infty$ functions $\kappa_{ij}$ defined as
	\begin{equation*}
	\kappa_{ij}(s) := 
	\begin{cases}
	\kappa_i(s), ~~~~& \text{if }i = j,\\
	\rho_{i}(\alpha_j^{-1}(s)), ~~~~& \text{if }i \neq j,
	\end{cases}
	\end{equation*}
	satisfy
	\begin{equation}\label{Assump: Kappa}
	\kappa_{i_1i_2}\circ\kappa_{i_2i_3}\circ\dots \circ \kappa_{i_{r-1}i_{r}}\circ\kappa_{i_{r}i_1} < \mathcal{I}_d,
	\end{equation}	
	for all sequences $(i_1,\dots,i_{r}) \in \{1,\dots,N\}^ {r}$ and $r\in \{1,\dots,N\}$. 
\end{assumption}
The small-gain condition~\eqref{Assump: Kappa} implies the existence of $\mathcal{K}_\infty$ functions $\varrho_i>0$ \cite[Theorem 5.5]{ruffer2010monotone}, satisfying
\begin{equation} \label{compositionality}
\max_{i,j}\Big\{\varrho_i^{-1}\circ\kappa_{ij}\circ\varrho_j\Big\} < \mathcal{I}_d, ~~~~i,j = \{1,\dots,N\}.
\end{equation}
\begin{remark}
	Note that \eqref{Assump: Kappa} is a standard small-gain assumption employed for investigating the stability of large-scale interconnected systems via ISS Lyapunov functions~\cite{dashkovskiy2007iss,dashkovskiy2010small}. This condition is automatically satisfied if each $\kappa_{ij}$ is less than identity (\emph{i.e.,} $\kappa_{ij}<\mathcal{I}_d, \forall i,j\in\{1,\dots,N\}$).
\end{remark}		
In the next theorem, we show that one can construct a CBC of $\mathfrak{S}$ using CSBC of $\mathfrak{S}_i$ if Assumption \ref{Assump: Gamma} holds and $\max_{i}\varrho_i^{-1}$ is concave (in order to employ Jensen's inequality \cite{Chandler1987IntroductionTM}). 

\begin{theorem}\label{Thm: Comp}
	Consider the interconnected dt-SCS
	$\mathfrak{S}=\mathcal{I}(\mathfrak{S}_1,\ldots,\mathfrak{S}_N)$ induced by $N\in\mathbb N_{\geq1}$ stochastic
	control subsystems~$\mathfrak{S}_i$. Suppose that each $\mathfrak{S}_i$ admits a CSBC $\mathds B_i$ as defined in Definition~\ref{csbc}. If Assumption~\ref{Assump: Gamma} holds and
	\begin{align}\label{compositionality1}
	\max_{i}\Big\{\varrho_i^{-1}(\beta_i)\Big\} > \max_{i}\Big\{\varrho_i^{-1}(\eta_i)\Big\},
	\end{align}
	then function $\mathds B(x)$ defined as
	\begin{equation}
	\label{Comp: Simulation Function}
	\mathds B(x) := \max_{i} \Big\{\varrho_i^{-1}(\mathds B_i(x_i))\Big\},
	\end{equation}
	is a CBC for the interconnected system 	$\mathfrak{S}=\mathcal{I}(\mathfrak{S}_1,\ldots,\mathfrak{S}_N)$ provided that $\max_{i}\varrho_i^{-1}$ for $\varrho_i$ as in \eqref{compositionality} is concave. 	
\end{theorem}

The proof of Theorem \ref{Thm: Comp} is provided in Appendix.

\begin{remark}
	Note that $\varrho_{i}$ in~\eqref{compositionality} plays a significant role in rescaling CSBC for subsystems while normalizing the effect of internal gains of other subsystems (cf. \cite{dashkovskiy2010small} for a similar argument but in the context of stability analysis via ISS Lyapunov functions). This rescaling issue mitigates the conservatism of condition~\eqref{compositionality1}, and hence, this condition is able to be satisfied in many scenarios (cf. case studies).
\end{remark}

So far, our discussion has been limited to providing probabilistic guarantees for safety properties for interconnected dt-SCS via control barrier certificates. In the next section, we introduce a more general class of specifications expressed  by deterministic finite automata, and thereafter, we provide a systematic method to obtain probabilistic guarantees for the satisfaction of such complex specifications.

\section{Specifications Expressed by DFA}
In this paper, we deal with a general class of specifications that can be expressed  by deterministic finite automata as the following definition.

\begin{definition}\label{DFA}
	A deterministic finite automaton (DFA) is a tuple $\mathcal{A}=(Q,q_0,\Sigma,\delta,F)$, where $Q$ is a finite set of states, $q_0 \in Q $ is the initial state, $\Sigma$ is a finite set of input symbols called alphabet, $\delta: Q \times \Sigma \rightarrow {Q}$ is the transition function and $F \subseteq Q$ represents the accepting states.  
\end{definition}

	We consider specifications that can be represented by accepting languages of DFA $\mathcal{A}$ as in Definition~\ref{DFA} over a set of atomic propositions $\mathcal{AP}$, \textit{i.e.,} $\Sigma=2^{\mathcal{AP}}$. For instance, all LTL specifications  over finite-time horizons (\textit{i.e.} LTL$_f$) \cite{synthesis_2015} can be represented by DFA which can be built using existing tools such as SPOT \cite{duret.16} and MONA \cite{henriksen_mona_1995}. Note that specifications represented by DFA are more expressive than LTL$_f$ \cite{ltlf}.

Let $\delta(q,\sigma)$ denote a state in the DFA that can be reached from state $q \in Q$ in the presence of a symbol $\sigma$. A finite word or trace $(\sigma_0, \sigma_1, \ldots , \sigma_{n-1}) \in \Sigma^{n}$ is accepted by the DFA if there exists a finite state run $\textbf{q} =(q_0, q_1, \ldots , q_n) \in Q^{n+1}$ such that $q_{i+1} = \delta(q_i, \sigma_i)$ for all $0 \leq i < n$ and $q_{n} \in F$. The accepting language of DFA $\mathcal{A}$ is denoted by $\mathcal{L}(\mathcal{A})$ which is the set of all finite words accepted by $\mathcal{A}$.  The complement of a DFA is simply acquired by interchanging its accepting and non-accepting states.

\begin{definition}
	For an interconnected dt-SCS $\mathfrak{S}=(X,U,\varsigma,f)$ and a DFA $\mathcal{A}=(Q,q_0,\mathcal{AP},\delta,F)$, consider a labeling function $L: X \rightarrow \mathcal{AP}$. For a finite state sequence $x_{M}=(x(0),x(1),\ldots,x(M-1)) \in X^{M}$ of a length $M \in \mathbb{N}$, the corresponding finite word over $\mathcal{AP}$ is given by $L(x) := (\sigma_0, \sigma_1,\ldots,\sigma_{M-1}) \in \mathcal{AP}^{M}$\!, where $\sigma_i=L(x(i))$ for all $i \in (0,1,\ldots,M-1)$.
\end{definition}

\begin{remark}
	A DFA $\mathcal{A}$ is normally constructed over the alphabet $\Sigma=2^{\mathcal{AP}}$. However, without loss of generality, we work here with the set of atomic propositions $\mathcal{AP}$ as the alphabet rather than its power set $2^{\mathcal{AP}}$, \textit{i.e.,} $\Sigma=\mathcal{AP}$. This is due to the fact that for any two atomic propositions $p_i,p_j \in \mathcal{AP}$, $i,j \leq \vert\mathcal{AP}\vert$, we have $p_i \wedge p_j = \emptyset$, and therefore edges with conjunctions between atomic propositions can be removed from the DFA. Moreover, other Boolean combinations like disjunction and negation can be easily resolved by adding parallel edges with simple atomic propositions in each of the edges.
\end{remark}

We now define the probability that solution processes of the interconnected system satisfy a specification over a time horizon $M$.  

\begin{definition}
	Consider an interconnected dt-SCS $\mathfrak{S}=(X,U,\varsigma,f)$, a specification given by the accepting language of a DFA $\mathcal{A}=(Q,q_0,\mathcal{AP},\delta,F)$ and a labeling function $L:X \rightarrow \mathcal{AP}$. Then, the  probability with which the solution process $x_M$ of $\mathfrak{S}$ of length $M \in \mathbb{N}$ started from an initial condition $x(0)=a$ under the controller $\nu(\cdot)$, satisfies the specification expressed by $\mathcal{A}$ is denoted by $\mathbb{P}^a_\nu\{L(x_M) \models \mathcal{A} \}$.
\end{definition}

The synthesis problem considered in this paper involves computing a controller in conjunction with a tight lower bound on the probability of satisfaction over the interconnected dt-SCS $\mathfrak{S}$. This problem can be formally presented as follows.

\begin{problem} \label{probstate}
	Given an interconnected dt-SCS $\mathfrak{S}=(X,U,\varsigma,f)$, a desired specification admitted by the accepting language of the DFA  $\mathcal{A}=(Q,q_0,\mathcal{AP},\delta,F)$ over a set of atomic propositions $\mathcal{AP}=\{p_0,p_1\ldots,p_R\}$,  $R \in \mathbb{N}$, and a labeling function $L:X \rightarrow \mathcal{AP}$, compute a controller $\nu(\cdot)$ and a constant $\varepsilon \in [0,1]$ such that $\mathbb{P}^a_\nu\{L(x_M) \models \mathcal{A} \} \geq \varepsilon$.
\end{problem}

To tackle this problem, we utilize a DFA representing the complement of the complex specification and decompose it into simpler reachability tasks. For each such task, we aim to find a suitable CBC as in Definition \ref{cbc} along with a controller for the interconnected dt-SCS. However, finding CBC for complex systems could be computationally expensive. In order to circumvent this complexity, we consider the interconnected dt-SCS $\mathfrak{S}=(X,U,\varsigma,f)$ as an interconnection of $N$ subsystems $\mathfrak{S_i}=(X_i,U_i,W_i,\varsigma_i,f_i,Y_i,h_i), i \in \{1,\ldots,N\}$, as explained in Section~\ref{interconnected}. 

Let $X_0$ and $X_u$ be two sets as introduced in Definition~\ref{cbc} that are connected to atomic propositions $\mathcal{AP}$ through some labeling function $L: X \rightarrow \mathcal{AP}$. We assume that those sets can be decomposed as $X_0= \prod_{i=1}^{N} X_{0_i}$ and $X_u=\prod_{i=1}^{N} X_{u_i}$. By doing so, one can simply compute a CSBC for each subsystem separately and utilize Theorem \ref{Thm: Comp} to obtain a CBC for the interconnected system.
Similarly, it is assumed that atomic propositions in the set $\mathcal{AP}$ can also be decomposed accordingly. This implies that sets $X_{0_i}$ and $X_{u_i}$, $i \in \{1,\ldots,N\}$, are also connected to the corresponding decomposed structure of $\mathcal{AP}$.       

In the following section, we discuss the procedure of sequential reachability decomposition. Later, we explain in detail the computation of probability bounds on the satisfaction of specifications.

\section{Sequential Reachability Decomposition}
In this subsection, we describe the sequential reachability decomposition, in which we divide a complex specification into simpler reachability tasks by utilizing the automaton representing the complement of the specification. This was initially proposed in~\cite{jagtap_formal_2019} but for a monolithic system. 

For a DFA $\mathcal{A} =(Q,q_0,\mathcal{AP},\delta,\bar{F})$  that describes the property of interest, consider the complement DFA $\mathcal{A}^{c}=(Q,q_0,\mathcal{AP},\delta,\bar{F})$ with $\bar{F}= Q \setminus F$ whose accepting language consists of all finite words not present in $\mathcal{L}(A)$.
A sequence $\textbf{q}=(q_0,q_1,\ldots,q_n) \in Q^{n+1}$ is an accepting state run of $\mathcal{A}^c$ if $q_n \in \bar{F}$ if there exists a finite word $\sigma(\textbf{q})=(\sigma_0, \sigma_1,\ldots,\sigma_{n-1})$ such that $q_{i+1} = \delta(q_i,\sigma_i)$ for all $i \in \{0,1,\ldots,n-1\}$. The length of the accepting state run is given by $|\textbf{q}|=n+1$. Let $Q_z \subseteq Q$ denote the set of states having self-loops. Let $\mathcal{R}_M, M \in \mathbb{N}$, be the set of all finite accepting state runs of at most length $M+1$ excluding self-loops, where
\begin{align*}
\mathcal{R}_M := &\{\textbf{q}=(q_0,q_1,\ldots,q_{m}) \in Q^{m+1} \,\big|\, m \leq M, q_m \in \bar{F}, q_{i} \neq q_{i+1}, \forall i <m\}.
\end{align*}
$\mathcal{R}_M$ can be computed algorithmically by considering the DFA as a directed graph $\mathcal{G}=(\mathcal{V},\mathcal{E})$, where $\mathcal{V}=Q$ are vertices representing states of the DFA and $\mathcal{E} \subseteq \mathcal{V} \times \mathcal{V}$ are edges such that $(q,q') \in \mathcal{E}$ if and only if $q' \neq q$, and there exists $\sigma \in \mathcal{AP}$ such that $\delta(q,\sigma)=q'$. It can be readily observed that a finite path starting at the vertex  $q_0$ and ending at a vertex $q_n \in \bar{F}$ is an accepting state run $\textbf{q}$ of $\mathcal{A}^c$ without any self-loop, and therefore, it belongs to $\mathcal{R}_M$. Using algorithms provided in the graph theory such as the depth-first search algorithm  \cite{russell_artificial_2003}, one can readily obtain $\mathcal{R}_M$. 

Now, for each $p \in \mathcal{AP}$, we define $\mathcal{R}^{p}_M$ as
$$\mathcal{R}^p_M := \{\textbf{q}= (q_0,q_1,\ldots,q_m) \in \mathcal{R}_M \,\big|\, \sigma(q_0,q_1)=p \in \mathcal{AP} \}.$$

We consider any $\textbf{q}=(q_0,q_1,\ldots,q_m) \in \mathcal{R}^p_M$ and define $\mathcal{P}^p(\textbf{q})$ as a set of all state runs augmented with a horizon as
\begin{equation} \label{eq:reach}
\mathcal{P}^p(\textbf{q}) := \{(q_l,q_{l+1},q_{l+2}), T_h(\textbf{q},q_{l+1}) | 0 \leq l \leq m-2 \},
\end{equation}
to decompose our specification into sequential reachabilities.
The horizon $T_h(\textbf{q},q_{l+1})=M+2-|\textbf{q}|$ for $q_{l+1} \in Q_z$, and $1$ otherwise. Consequently, we define $\mathcal{P}_{M}(\mathcal{A}^c)= \bigcup_{p \in \mathcal{AP}} \bigcup_{\textbf{q} \in \mathcal{R}^p_M}\mathcal{P}^p(\textbf{q})$ as the set of all reachability elements arising from different accepting state runs of a length less than or equal to $M+1$.

\begin{remark}\label{remarkdecomp}
	Note that $\mathcal{P}^p(\textbf{q})=\emptyset$ for those accepting state runs whose length is $2$. Any such sequences begin from a subset of the state space that already violates the desired specification and the outcome is accordingly a trivial zero probability for satisfaction of the specification. Hence, we neglect such accepting state runs.
\end{remark}

\begin{remark}
	The self-loops play a pivotal role in the computation of the time horizon $T_h(\textbf{q},q_{l+1})$ for any reachability element $\vartheta\!=\!(q_l, q_{l+1}, q_{l+2})$ $\!\in\! \mathcal{P}^p(\textbf{q})$. This is crucial to account for the number of time steps that the solution process can remain in the self-loop $q_{l+1} \in Q_z$ before reaching $q_{l+2}$ \cite{jagtap_formal_2019}.
\end{remark}

We illustrate the procedure of decomposition into sequential reachabilities with the help of a running example.

\begin{figure} 
	\begin{center}
		\includegraphics[scale=0.24]{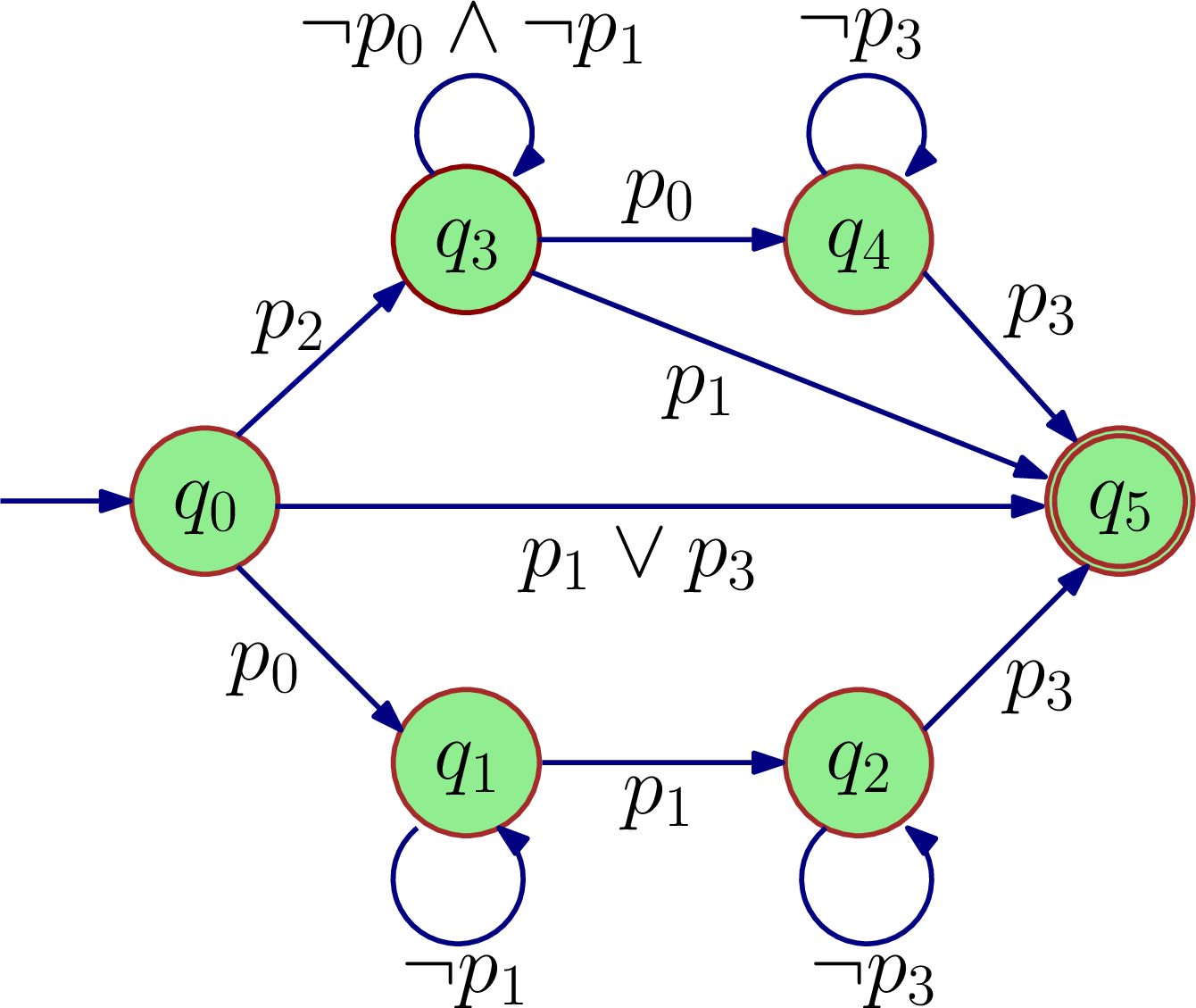} 
		\caption{DFA $\mathcal{A}^c$ employed in Example~\ref{example}.}
		\label{fig:DFAexample}
	\end{center}
\end{figure}

\begin{example} \label{example}
	Consider a DFA $\mathcal{A}^c$ as shown in Fig.~\ref{fig:DFAexample}. According to the definition of DFA, initial state is $q_0$, set of atomic propositions $\mathcal{AP}=\{p_0,p_1,p_2,p_3\}$ and set of final states $\bar{F}=\{q_5\}$. The set of states with self-loops are given by $Q_z= \{q_1, q_2, q_3, q_4\}$. We only consider accepting state runs with lengths less than or equal to $5$, \textit{i.e.}, $M=4$. The set of such accepting state runs without self-loops is 
	\begin{equation*}
	\mathcal{R}_4=\{(q_0,q_5), (q_0,q_3,q_5), (q_0, q_1, q_2, q_5), (q_0, q_3, q_4, q_5) \}.
	\end{equation*}
	
	The sets $\mathcal{R}^p_4$ for all $p \in \mathcal{AP}$ are given by
	\begin{align*}
	&\mathcal{R}^{p_0}_4=\{(q_0,q_1,q_2,q_5)\}, \ \  \mathcal{R}^{p_1}_4=\{(q_0,q_5)\}, \\ &\mathcal{R}^{p_2}_4=\{(q_0,q_3,q_5),(q_0,q_3,q_4,q_5)\}, \ \ \mathcal{R}^{p_3}_4=\{(q_0,q_5)\}. 
	\end{align*}
	
	For all $\textbf{q} \in \mathcal{R}^p_4$, we define $\mathcal{P}^p(\textbf{q})$ as
	\begin{align*}
	&\mathcal{P}^{p_0}(q_0,q_1,q_2,q_5)= \{(q_0,q_1,q_2, 2), (q_1,q_2,q_5,2)\}, ~~\mathcal{P}^{p_1}(q_0,q_5)=\mathcal{P}^{p_3}(q_0,q_5)=\emptyset, \\
	&\mathcal{P}^{p_2}(q_0,q_3,q_5)=\{(q_0,q_3,q_5,3)\}, ~~\mathcal{P}^{p_2}(q_0,q_3,q_4,q_5)=\{(q_0,q_3,q_4,2),(q_3,q_4,q_5,2)\}.
	\end{align*} 
	For each $\textbf{q} \in \mathcal{R}_4$, the corresponding finite words $\sigma(\textbf{q})$ are given by
	\begin{align*}
	&\sigma(q_0,q_5)=\{(p_1 \vee p_3)\}, ~~\sigma(q_0,q_3,q_5)=\{(p_2,p_1)\}, \\
	&\sigma(q_0,q_1,q_2,q_5)=\{(p_0,p_1,p_3)\}, \\ &\sigma(q_0,q_3,q_4,q_5)=\{(p_2,p_0,p_3)\}. \\
	\end{align*}
\end{example}

Now, for each reachability task, we construct an appropriate CBC along with a corresponding controller to obtain an upper bound on the probability that the interconnected system $\mathfrak{S}$ reaches unsafe regions in finite-time horizons. We now raise the following lemma to compute CBCs and reachability probabilities.      

\begin{lemma} \label{reachlem}
	For an accepting state run $\textbf{q} \in \mathcal{R}^p_M$  for some $M \in \mathbb{N}$ and some $p \in \mathcal{AP}$, consider the reachability element $\vartheta=(q,q',q'',T_h) \in \mathcal{P}^p(\textbf{q})$. If there exists a CBC and a controller $\nu(\cdot)$ such that conditions \eqref{sys2}-\eqref{cbceq} hold with $X_0 = L^{-1}(\sigma(q,q'))$ and $X_u= L^{-1}(\sigma(q',q''))$, then the upper bound on the probability that a solution process of dt-SCS $\mathfrak{S}$ starts from an initial state $a \in X_0$ under the controller $\nu(\cdot)$ and reaches $X_u$ within the finite-time horizon $[0,T_h)$ is obtained from \eqref{eqlemma2} and is denoted by $\varkappa_{\vartheta T_h}$.
\end{lemma}

Once we compute the CBCs and the corresponding probabilities for all individual reachability elements, we combine them to obtain an upper bound probability of satisfaction of the property expressed by $\mathcal{A}^c$, or in other words, an upper bound on the probability of violation of the specification given by the accepting language of $\mathcal{A}$. Consequently, we quantify a lower bound on the probability of satisfaction together with a controller that ensures satisfaction of the desired specification. The next section explains the structure of this controller as well as the proposed procedure to compute the lower bound on the probability that the overall complex specification is satisfied by the interconnected system.

\section{Controller and Probability Computation}

Ideally, one has to compute the CBC and a suitable controller for each element of $\mathcal{P}_M(\mathcal{A}^c)$. However, it is ambiguous when utilizing the controller in the closed loop at those states of automaton where there is more than one edge emanating from the state. Therefore, we combine different reachability tasks into a single partition set and present the controller for the interconnected system as a switching one. This is explained in the following subsection. Later, we discuss the computation of the lower bound on the probability that the interconnected system satisfies the complex specification represented by the accepting language of DFA $\mathcal{A}$.

\subsection{Controller Structure} \label{subcontrolpolicy}

Computing CBC and its corresponding controller for the specification described by each individual reachability element could be ambiguous once applying the controllers in a closed loop fashion. To clarify this, we consider the DFA $\mathcal{A}^c$ from Fig.~\ref{fig:DFAexample}. $\mathcal{P}_M(\mathcal{A}^c)$ is the set of all reachability elements for accepting state runs of a length of at most $M+1$, as obtained in Example \ref{example}. The elements $\vartheta_1=(q_0,q_3,q_4,2)$ and $\vartheta_2=(q_0,q_3,q_5,3)$ constitute two individual problems: one for computing the upper bound of reaching the region $L^{-1}(p_0)$ from $L^{-1}(p_2)$ and the other one for reaching the region $L^{-1}(p_1)$ from the same region $L^{-1}(p_2)$. Ideally, one should find two different CBCs and controllers. But since there are two outgoing transitions from state $q_3$, namely $\delta(q_3,p_0)$ and $\delta(q_3,p_1)$, computing different controllers means that the region $L^{-1}(p_2)$ employs two different controllers simultaneously and this issue results in ambiguity in the closed-loop system. 

One potential solution to tackle this problem is to replace $X_u$ in Lemma \ref{reachlem} with the union of regions  and combine the two reachability problems into one. This results in a common CBC and controller for different reachability elements. In other words, we partition $\mathcal{P}_M(\mathcal{A}^c)$ and combine the reachability elements with the same CBC and controller and place them in a single partition set. Consequently, we obtain a switching controller since multiple locations in the automaton $\mathcal{A}^c$ admitting different controllers. In order to represent such a switching policy, a DFA $\mathcal{A}^c_s$ is constructed. This procedure has been adapted from~\cite{jagtap_formal_2019}. 

As discussed before, reachability elements admitting a common CBC and controller are combined together in a single partition set. Such sets can be defined as 
\begin{align*}
&\gamma_{(q,q',\Delta(q'))} := \{(q,q',q'',T) \in \mathcal{P}_M(\mathcal{A}^c)\,\big|\, q,q',q'' \in Q \text{ and } q'' \in \Delta(q')\},
\end{align*}
where for any state $q \in Q$, $\Delta(q)$ is the set of states that can be reached from $q$ in one transition.

For each partition set $\gamma_{(q,q',\Delta(q'))}$, we denote its corresponding CBC and controller as $\mathds{B}_{\gamma_{(q,q',\Delta(q'))}}(x)$ and $\nu_{\gamma_{(q,q',\Delta(q'))}}$, respectively. For all reachability elements $\vartheta \in \mathcal{P}_M({\mathcal{A}^c})$, we therefore have
\begin{align} \nonumber
&\mathds{B}_\vartheta(x)=\mathds{B}_{\gamma_{(q,q',\Delta(q'))}}(x) \text{ and } \nu_\vartheta(x)=\nu_{\gamma_{(q,q',\Delta(q'))}}(x),
\text{ if } \vartheta \in \gamma_{(q,q',\Delta(q'))}. \nonumber
\end{align}
This results in a switching controller, where multiple locations on the automaton dictate different controllers. For the DFA $\mathcal{A}^c=(Q,q_0,\mathcal{AP},\delta,\bar F)$ with $\bar F = Q \backslash F$, the corresponding DFA representing switching mechanism is given by DFA $\mathcal{A}^c_s=(Q_{s},q_{0s},\mathcal{AP}_s,\delta_s,F_s)$ where
$Q_s := q_{0s} \cup \{(q,q',\Delta(q')) \ | \ q,q' \in Q\backslash \bar F\} \cup \bar F$ is the set of states, $q_{0s}:= (q_0,\Delta(q_0))$ is the initial state, $\mathcal{AP}_{s}=\mathcal{AP}$ is the set of atomic propositions and $F_{s}=\bar F$ is the set of final states. The transition function $\delta_{s}$ is defined as

\begin{itemize}
	\item for $q_{0s}=(q_0,\Delta(q_0))$, 
	\begin{itemize}
		\item $\delta_s((q_0,\Delta(q_0)),\sigma_{(q_0,q_0')})=(q_0,q_0',\Delta(q_0'))$ where $q_0' \in \Delta(q_0)$,
	\end{itemize}
	\item for all $q_s=(q,q',\Delta(q')) \in Q_s \backslash (q_{0s} \cup \bar F) $,
	\begin{itemize}
		\item $\delta_s((q,q',\Delta(q')), \sigma_{(q',q'')})=(q',q'',\Delta(q''))$, where $q,q',q'' \in Q, q'' \in \Delta(q') \text{ and } q'' \notin \bar F$, and
		\item $\delta_s((q,q',\Delta(q')),\sigma_{(q',q'')})=q''$ where $q,q',q'' \in Q$, $q'' \in \Delta(q')$ and $q'' \in \bar F$.  
	\end{itemize}
\end{itemize}

Now, the controller for Problem \ref{probstate} is formally given by
\begin{equation} \label{eq:controlpolicy}
\nu(x,q_s)=\nu_{(\gamma_{q_s,L(x),q'_s})}(x), \ \ \forall(q_s,L(x),q'_s) \in \delta_s.
\end{equation}

\begin{remark}
	The switching control policy is a Markov policy on the augmented space of $X \times Q_s$. It is a history dependent policy over the state $X$ of the system as mentioned in Definition \ref{controlpolicy}.
\end{remark}

\def\example{\par{\textit{Example 5.8}} \ignorespaces}
\def\endexaple{}

\begin{figure}
	\center
	\includegraphics[scale=0.75]{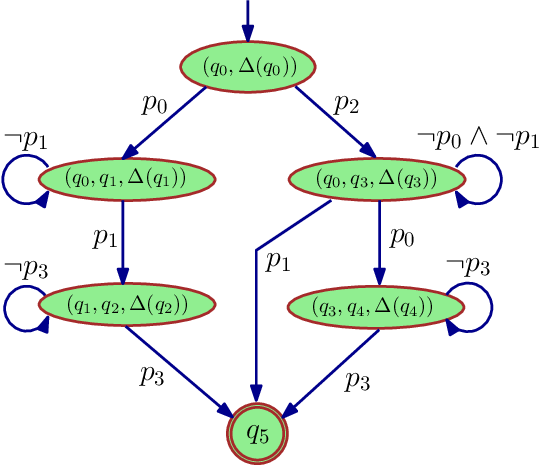} 
	\caption{DFA $\mathcal{A}^c_s$ representing the switching mechanism.}
	\label{fig:DFAswitch}
\end{figure}

{\bf Example~\ref{example} (continued.)} 
The DFA $\mathcal{A}^c_s$ representing the switching mechanism between policies for Example \ref{example} is shown in Fig.~\ref{fig:DFAswitch}.	

\subsection{Probability Computation}

For each individual reachability element given by $\vartheta=(q,q',q'',T_h) \in \mathcal{P}_M(\mathcal{A}^c)$, we first compute upper bounds on reachability probabilities and then combine them to provide an upper bound on the probability that the specification represented by the language of DFA $\mathcal{A}$ is violated, which is provided by the following theorem. 

\begin{theorem}\label{th:sumprod}
	For a specification given by the accepting language of DFA $\mathcal{A}$, let $\mathcal{A}^c$ represent the complement of $\mathcal{A}$. For $\mathcal{A}^c$, let $\mathcal{R}^{p}_M$ be the set of all accepting state runs of the length of at most $M+1$  and $\mathcal{P}^p(\textbf{q})$ be the set of state runs of length $3$ augmented with the horizon $T_h$ for $p \in \mathcal{AP}$. Then the probability that the solution processes of dt-SCS $\mathfrak{S}$ starting from any initial state $a \in L^{-1}(p)$ satisfy the specification represented by $\mathcal{A}^c$ under the controller in  \eqref{eq:controlpolicy} within the time horizon $[0,M) \subseteq \mathbb{N}$ is upper bounded by
	
	\begin{align}\label{eq:sumprod}
	&\mathbb{P}^a_\nu\{L(x_M) \models \mathcal{A}^c\}  \leq  \sum_{\textbf{q} \in \mathcal{R}^p_M}  \prod_{\vartheta \in \mathcal{P}^p(\textbf{q})}\{\varkappa_{\vartheta T_h} | \vartheta=(q,q',q'',T_h) \in  \mathcal{P}^p(\textbf{q})\}, 
	\end{align}	
	where $\varkappa_{\vartheta T_h}$ is obtained using Lemma \ref{reachlem} and is the upper bound on the probability that solution processes of the system $\mathfrak{S}$ start from $X_0 := L^{-1}(\sigma(q,q'))$ and reach $X_u := L^{-1}(\sigma(q',q''))$ within the time horizon $[0,T_h) \subseteq \mathbb{N}$.  
	
\end{theorem}   

The proof of Theorem~\ref{th:sumprod} is provided in Appendix.

\begin{remark}
	If no CBC is found for a certain element $\vartheta \in \mathcal{P}^p(\textbf{q})$, the corresponding probability $\varkappa_{\vartheta T_h}$ for that element  should be replaced by a trivial probability bound $1$ in  \eqref{eq:sumprod}. To obtain a \emph{non-trivial} probability of satisfaction, CBC should be found for at least one element.
\end{remark}

\begin{remark}
	The proposed bounds in~\eqref{eq:sumprod} can be improved by minimizing $\eta,c$ for some fixed $\beta$, for each reachability element $\vartheta$. Since CBC of $\mathfrak{S}$ is obtained compositionally via~\eqref{Comp: Simulation Function}, these parameters depend on $\eta_i$, $c_i$ for some fixed $\beta_i$, for all subsystems $\mathfrak{S}_i$, $i \in \{1, \ldots, N\}$. The bound is then improved by minimizing $\eta_i,c_i$ for all $i$, via the bisection method~\cite{jagtap_formal_2019}. Note that equation~\eqref{Comp: Simulation Function} allows for CSBCs of some subsystems to compensate the undesirable parameters of CSBCs of other subsystems as long as condition~\eqref{compositionality1} holds. 
\end{remark}

In the following corollary, we provide the formula for computing the lower bound on the probability that the interconnected system $\mathfrak{S}$ satisfies the desired specification represented by the DFA $\mathcal{A}$.

\begin{corollary} \label{cor:mainprob}
	The probability that the solution processes of $\mathfrak{S}$ start from any initial state $a \in L^{-1}(p)$ and satisfy the specification given by the accepting language of DFA $\mathcal{A}$ over a finite-time horizon $[0,M) \subseteq \mathbb{N}$ is lower bounded by
	
	\begin{align}
	&\mathbb{P}^a_\nu\{L(x_M) \models \mathcal{A} \}  \geq  1 -  \sum_{\textbf{q} \in \mathcal{R}^p_M}  \prod_{\vartheta \in \mathcal{P}^p(\textbf{q})} \hspace{-0.5em}\{\varkappa_{\vartheta T_h} | \vartheta  =  (q,q',q'',  T_h) \in  \mathcal{P}^p(\textbf{q}) \}. 
	\end{align}
\end{corollary}

Thus far, we elaborated on the procedure of dividing a complex specification into simpler specifications, computing an upper bound on the probability of reachability for those simpler tasks by finding a CBC and suitable controller for interconnected dt-SCS, and finally combining them in a sum-product form in order to obtain a tight lower bound on the probability of satisfactions. It has already been discussed that finding a CBC for a large-scale interconnected dt-SCS is computationally very expensive. Results from Theorem~\ref{Thm: Comp} can be employed to obtain  a CBC and a controller for the interconnected system. In the next section, we provide two systematic approaches to search for CSBC and their corresponding controllers.

\section{Computation of CSBC and Corresponding Controllers} \label{computation}

In  this section, we provide suitable methods to search for CSBC and synthesize corresponding controllers satisfying simple safety specifications decomposed from DFA $\mathcal{A}^c$ for subsystems $\mathfrak{S}_i$. We propose two different approaches: one is based on the sum-of-squares (SOS) optimization problem and another one relies on counter-example guided inductive synthesis (CEGIS) framework.

\subsection{Sum-of-Squares Optimization Problem} \label{sossubsec}

Here, we reformulate conditions~\eqref{subsys1}-\eqref{csbceq} as an SOS optimization problem~\cite{Parrilo2003}, where CSBC is restricted to be a non-negative polynomial that can be written as a sum of squares of different polynomials. To do so, we need to raise the following assumption.

\begin{assumption} \label{assumeSOS}
	The stochastic control subsystem $\mathfrak{S}_i$ has a continuous state set $X_i \subseteq \mathbb{R}^{n_i}$, and continuous external and internal input sets $U_i \subseteq \mathbb{R}^{m_i}$ and $W_i \subseteq \mathbb{R}^{p_i}$. Its vector field $f_i: X_{i} \times U_i \times W_i \times \mathcal V_{\varsigma_i} \rightarrow X_i$ is a polynomial function of the state $x_i$, the external input $\nu_i$, and the internal input $w_i$. We also assume that the output map $h_i:X_i\rightarrow Y_i$ and $\mathcal{K}_{\infty}$ functions $\alpha_i$ and $\rho_{i}$ are polynomial.
\end{assumption}

Under Assumption \ref{assumeSOS}, one can reformulate conditions~\eqref{subsys1}-\eqref{csbceq} as an SOS optimization problem to search for a polynomial CSBC $\mathds B_i$ and a polynomial controller $\nu_i(\cdot)$ for the subsystem $\mathfrak{S}_i$. Correspondingly, one can utilize the compositionality results of the paper to construct CBCs and controllers for interconnected systems. The following lemma provides the SOS formulation. 

\begin{lemma}\label{sos}
	Suppose Assumption \ref{assumeSOS} holds and sets $X_i$ , $X_{0_i}$, $X_{u_i}$ can be defined by vectors of polynomial inequalities $X_i=\{x_i \in \mathbb{R}^{n_i} \mid g_i(x_i) \geq 0\}$, $X_{0_i}=\{x_i \in \mathbb{R}^{n_i} \mid g_{0_i}(x_i) \geq 0\}$, and $X_{u_i}=\{x_i \in \mathbb{R}^{n_i} \mid g_{u_i}(x_i) \geq 0\}$, where the inequalities are provided element-wise. Similarly, let external and internal input sets $U_i$ and $W_i$ be defined by vectors of polynomial inequalities $U_i=\{\nu_i\in\R^{m_i}\mid g_{\nu_i}(\nu_i)\geq0\}$ and $W_i=\{w_i\in\R^{p_i}\mid g_{w_i}(w_i)\geq0\}$. Suppose for a given control subsystem $\mathfrak{S}_i$, there exists a sum-of-squares polynomial $\mathds B_i(x_i)$, constants $\eta_i,\bar c_i \in \R_{\geq 0}$, $\beta_i \in \R_{> 0}$, functions $\bar \rho_i\in\mathcal{K}_\infty\cup\{0\}$, $\alpha_i,\bar \kappa_i\in\mathcal{K}_\infty$, with $ \bar \kappa_i < \mathcal{I}_d$, vectors of sum-of-squares polynomials $\lambda_{0_i}(x_i), \lambda_{u_i}(x_i)$, $\lambda_i(x_i)$, $\hat\lambda_i(x_i,\nu_i,w_i)$, $\lambda_{\nu_i}(x_i,\nu_i,w_i)$, $\lambda_{w_i}(x_i,\nu_i,w_i)$ and polynomials $\lambda_{\nu_{ji}}(x_i)$ corresponding to the $j^{\text{th}}$ input in $\nu_i=(\nu_{1i},\ldots,\nu_{m_{i}}) \in U_i \subseteq \mathbb{R}^{m_i}$ of appropriate dimensions such that the following expressions are sum-of-squares polynomials:
	\begin{align} \label{sos1}
	&\mathds B_i(x_i)-\lambda^{T}_i(x_i)g_i(x_i)-\alpha_i(h_i^T(x_i)h_i(x_i)),\\\label{sos2}
	-&\mathds B_i(x_i)-\lambda^{T}_{0_i}(x_i)g_{0_i}(x_i)+\eta_i,\\\label{sos3}
	&\mathds B_i(x_i)-\lambda^{T}_{u_i}(x_i)g_{u_i}(x_i)-\beta_i,\\\notag
	-&\mathbb{E}\Big[\mathds B_i(f_i(x_i,\nu_i,w_i,\varsigma_i)) \mid x_i,\nu_i,w_i\Big]+\bar\kappa_i(\mathds B_i(x_i))+ \bar\rho_{i}(\frac{w_i^Tw_i}{p_i})+\bar c_i \notag \\&~~~-\sum_{j=1}^{m_i}(\nu_{ji}-\lambda_{\nu_{ji}}(x_i))-\hat\lambda^{T}_i(x_i,\nu_i,w_i)g_i(x_i)-\lambda_{\nu_i}^{T}(x_i,\nu_i,w_i)g_{\nu_i}(\nu_i)-\lambda_{w_i}^{T}(x_i,\nu_i,w_i)g_{w_i}(w_i), \label{sos4}
	\end{align}
	where $p_i$ is the dimension of the internal input $w_i$. Then $\mathds B_i(x_i)$ is a CSBC satisfying conditions~\eqref{subsys1}-\eqref{csbceq} and $\nu_i=[\lambda_{\nu_{1i}}(x_i);\dots;\lambda_{\nu_{m_{i}}}(x_i)],~ i\in\{1,\dots,N\}$, is the corresponding controller for the subsystem $\mathfrak{S}_i$. The parameters satisfying the conditions are given by
	\begin{align}\notag
	&{\kappa}_i =\mathcal{I}_d -(\mathcal{I}_d - {\pi}_i )\circ (\mathcal{I}_d - \bar\kappa_i),\\\notag
	&\rho_i=(\mathcal{I}_d + \bar {\delta}_i)\circ(\mathcal{I}_d - \bar\kappa_i )^{-1}\circ {\pi}_i ^{-1}\circ\bar\pi_i\circ\bar \rho_i,\\\notag
	&c_i=(\mathcal{I}_d + \bar {\delta}_i^{-1})\circ(\mathcal{I}_d - \bar\kappa_i)^{-1}\circ {\pi}_i  ^{-1}\circ\bar\pi_i\circ(\bar \pi_i - \mathcal{I}_d)^{-1} ( \bar c_i),
	\end{align}
	where $\bar {\delta}_i, {\pi}_i ,\bar\pi_i$ are some arbitrarily chosen $\mathcal{K}_\infty$ functions so that $\mathcal{I}_d - {\pi}_i  \in \mathcal{K}_\infty$ and $\bar\pi_i - \mathcal{I}_d \in \mathcal{K}_\infty$.	
	
	The proof of Lemma~\ref{sos} is provided in Appendix.
	
\end{lemma}  
\begin{remark}
		Inequalities \eqref{subsys1} and \eqref{csbceq} consider infinity norms over $h_i(x_i)$ and $w_i$, respectively. Since such norms cannot be expressed as polynomials, we convert infinity norms to Euclidean ones by applying their corresponding weights in order to present expressions \eqref{sos1} and \eqref{sos4} as polynomials.
\end{remark}

\begin{remark}
	Since $\bar\kappa_i(\cdot)$ in \eqref{sos4} can cause nonlinearity on unknown parameters of $\mathds B_i$, one can consider a linear function $\bar\kappa_i(s)=\hat\kappa_i s, \forall s\in\R_{\geq 0}$, with some given constant $\hat\kappa_i\in\R_{> 0}$ to handle this nonlinearity.
\end{remark}

\begin{remark}
	Note that Lemma~\ref{sos} is different from~\cite[Lemma 5.6]{jagtap_formal_2019} in two senses. First, we have here an additional condition~\eqref{sos1} which is required in networks of control systems for the satisfaction of compositionality conditions. In addition, the condition \eqref{sos4} consists of an internal input term which also plays a role in the compositionality results.
\end{remark}

\subsection{Counter-Example Guided Inductive Synthesis}

In this approach, one can find a CSBC of a given parametric form, \emph{e.g.,} polynomials, by utilizing satisfiability modulo theories (SMT) solvers such as Z3 \cite{de2008z3}, dReal \cite{gao_-complete_2012} or MathSat \cite{cimatti_mathsat5_2013}. This framework does not require any restrictions on the underlying dynamics and is applicable under the following assumption.

\begin{assumption} \label{assump:cegis}
	Each subsystem $\mathfrak{S}_i,\ i \in \{1,\ldots,N\}$, has a compact state set $X_i$, a compact internal input set $W_i$ and a compact external input set $U_i$.
\end{assumption}

Under Assumption \ref{assump:cegis}, we propose the following lemma to reformulate conditions \eqref{subsys1}-\eqref{csbceq} as a satisfiability problem.
\begin{lemma}
	Consider a stochastic control subsystem $\mathfrak{S}_i=(X_i,U_i,W_i,\varsigma_i,f_i,Y_i,h_i), i \in \{1,\ldots,N\}$, satisfying Assumption \ref{assump:cegis}. Suppose there exist a function $\mathds{B}_i(x_i)$, constants $\eta_i,c_i \in \mathbb{R}_{\geq 0},  \beta_i \in \mathbb{R}_{>0}$, functions $\rho_i \in \mathcal{K}_\infty \cup \{0\}, \alpha_i, \kappa_i \in \mathcal{K}_\infty$, with $\kappa_i < \mathcal{I}_d$ such that 
	
	\begin{align*} 
	&\bigwedge_{x_i \in X_i}  \hspace{-0.5em} \!\!\big(\mathds{B}_i(x_i) \hspace{-0.2em} \geq \hspace{-0.2em} \alpha_i(\|h_i(x_i)\|^2)\big)\!\! \hspace{-0.5em} \bigwedge_{x_i \in X_{0_i}}\!\!  \hspace{-0.5em} \big(\mathds{B}_i(x_i)  \hspace{-0.2em} \leq  \hspace{-0.2em} \eta_i \big) \!\!
	\hspace{-0.5em} \bigwedge_{x_i \in X_{u_i}}\!\!  \hspace{-0.5em} \big(\mathds{B}_i(x_i) \hspace{-0.2em} \geq \hspace{-0.2em} \beta_i\big) \!\!\bigwedge_{x_i \in X_i}\bigvee_{\nu_i \in U_i} \bigwedge_{w_i \in W_i} \!\!\Big(\mathbb{E}\Big[\mathds{B}_i(f_i(x_i,\nu_i,w_i,\varsigma_i)) \,\big|\, x_i, \nu_i, w_i\Big] \nonumber \\
	&\hspace{25.7em} \leq \max\{{\kappa}_i(\mathds{B}_i(x_i)),\rho_i(\|w_i\|^2),c_i \}\Big).
	\end{align*}
	
	Then $\mathds{B}_i(x_i)$ is a CSBC satisfying conditions \eqref{subsys1}-\eqref{csbceq}. 
\end{lemma}

\section{Case Studies}

In this section, we demonstrate our proposed results using two physical case studies. The first one is a room temperature regulation in a circular building with $1000$ rooms, where we compositionally synthesize controllers for regulating the temperature in each room. The second one is a \emph{fully-interconnected} Kuramoto network with $100$ \emph{nonlinear} oscillators where we synthesize hybrid controllers to ensure satisfaction of a complex specification given by a deterministic finite automaton.     

\subsection{Room Temperature Network} \label{example1}

We first apply our approaches to a room temperature network in a circular building. The model of this case study is borrowed from~\cite{Meyer.2018} by including stochasticity as an additive noise. The evolution of the temperature $T(\cdot)$ in the interconnected system is governed by the following dynamics
\begin{equation*}
\mathfrak{S}: 
T(k+1)=AT(k) + \mu T_H\nu(k) + \iota T_E + 0.1\varsigma(k),
\end{equation*}
where $A \in \mathbb{R}^{n \times n}$ is a matrix with diagonal elements given by $\bar a_{ii}=(1-2 \epsilon-\iota-\mu\nu_i(k))$, off-diagonal elements $\bar a_{i,i+1}=\bar a_{i+1,i}=\bar a_{1,n}=\bar a_{n,1}= \epsilon$, $i\in \{1,\ldots,n-1\}$, and all other elements are identically zero. Parameters $\epsilon = 0.005$, $\iota = 0.06$, and $\mu = 0.145$ are conduction factors between rooms $i \pm 1$ and $i$, the external environment and the room $i$, and the heater and the room $i$, respectively.  Outside temperatures are the same for all rooms: $T_{ei}=-15\,{}^{\circ}\mathsf{C}$, $\forall i\in\{1,\ldots,n\}$, and the heater temperature is $T_H=45\,{}^{\circ}\mathsf{C}$. Moreover, $ T(k)=[T_1(k);\ldots;T_n(k)]$, $\varsigma=[\varsigma_1(k);\ldots;\varsigma_n(k)]$, $\nu(k)=[\nu_1(k);\ldots;\nu_n(k)]$, and $T_E=[T_{e_1};\ldots;T_{e_n}]$.

\begin{figure}
	\center
	\includegraphics[width=0.34\linewidth]{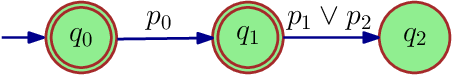}
	\caption{DFA $\mathcal{A}$ representing the specification.}
	\label{temp_spec}
\end{figure}

\begin{figure}
	\center
	\includegraphics[width=0.34\linewidth]{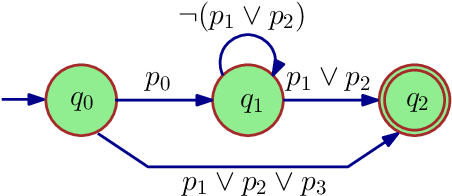}
	\caption{DFA $\mathcal{A}^c$ representing the complement of the specification.}
	\label{temp_spec_comp}
\end{figure}

We consider the regions of interest as $X^0=[19.5,20]^n, X^1=[1,17]^n, X^2=[23,50]^n$ and $X^3=T \backslash (X^0 \cup X^1 \cup X^2)$. Each region is associated with an element of atomic propositions given by $\mathcal{AP}=\{p_0,p_1,p_2,p_3\}$ through a labeling function $L: X \rightarrow \mathcal{AP}$ so that $L(x_l)=p_l, \ \forall x_l \in X^l, l \in \{0,1,2,3\}$. The main goal is to synthesize a controller for the system $\mathfrak{S}$ such that the solution processes of the system satisfy the language specified by DFA $\mathcal{A}$ shown in Fig.~\ref{temp_spec} for a finite-time horizon $T_d=10$. This means that temperature of all rooms are safely maintained between $17^{\circ}\mathsf{C}$ and $23^{\circ}\mathsf{C}$ for all time $k \in [0,T_d)$, with an initial temperature between $19.5^{\circ}\mathsf{C}$ and $20^{\circ}\mathsf{C}$. The complement of the DFA $\mathcal{A}$ is represented by  $\mathcal{A}^c$ as shown in Fig.~\ref{temp_spec_comp}. 

We consider only the set of accepting state runs $\mathcal{R}^p_M$ without self-loops for each $p \in \mathcal{AP}$ in the DFA $\mathcal{A}^c$ with $M=10$. Therefore, we have $\mathcal{R}^{p_0}_M = \{(q_0,q_1,q_2)\}$ and $\mathcal{R}^{p_1}_M=\mathcal{R}^{p_2}_M=\mathcal{R}^{p_3}_M=\{(q_0,q_2)\}$. We now first divide our specification into reachability tasks by identifying $\mathcal{P}^p(\textbf{q})$ for all $p \in \mathcal{AP}$. We have $\mathcal{P}^{p_1}(q_0,q_2)=\mathcal{P}^{p_2}(q_0,q_2)=\mathcal{P}^{p_3}(q_0,q_2)=\emptyset$ as $\textbf{q}=(q_0,q_2)$ is only an accepting state run of a length $2$ and admits a trivial probability bound of 1 (cf. Remark \ref{remarkdecomp}). The set $\mathcal{P}^{p_0}(\textbf{q})=\{(q_0,q_1,q_2,9)\}$ is obtained from \eqref{eq:reach}. As it can be seen, we have only one reachability element $\vartheta=(q_0,q_1,q_2,9)$ for which we need to find the existence of a suitable control barrier certificate and a corresponding controller along with an upper bound on the probability that the region corresponding to $L^{-1}(\sigma(q_1,q_2))$ is reached. To do so, we now consider our network $\mathfrak{S}$ as an interconnection of $n=1000$ subsystems (individual rooms) $\mathfrak{S}_i$ represented by
\begin{align}\notag
\mathfrak{S}_i: \begin{cases}
T_i(k+1) = \bar aT_i(k) + \mu T_H \nu_i(k) +  \epsilon w_i(k) + \iota T_{ei} + 0.1\varsigma_i(k), \\
y_i(k)=T_i(k).
\end{cases}
\end{align}

One can readily verify that $\mathfrak{S}=\mathcal{I}(\mathfrak{S}_1,\ldots,\mathfrak{S}_n)$ where $w_i(k) = [T_{i-1}(k);T_{i+1}(k)]$ (with $T_0 = T_n$ and $T_{n+1} = T_1$). We utilize the software tool \textsf{SOSTOOLS}~\cite{papachristodoulou2013sostools} and the SDP solver \textsf{SeDuMi}~\cite{sturm1999using} to compute a CSBC as described in Section~\ref{sossubsec}. Based on Lemma~\ref{sos}, we compute
a CSBC of an order $2$ as $\mathds B_i(T_i) = 0.7659T_i^2-30.24T_i+298.5$ and the corresponding controller of
an order $1$ as $\nu_i(T_i) = -0.012T_i+0.8$, $\forall i\in\{1,\dots,n\}$. Furthermore, the corresponding constants and functions in Definition~\ref{csbc} satisfying conditions~\eqref{subsys1}-\eqref{csbceq} are computed as $\eta_i = 0.13, \beta_i = 4.4, c_i = 0.0139, \alpha_i (s) = 5 \times 10^{-5}s, \kappa_i (s) = 0.99s,$ and  $\rho_i(s) = 4.99 \times 10^{-5}s, \forall s\in\R_{\geq 0}$.

In order to construct a CBC for the interconnected system using CSBC of subsystems,  we now check the small-gain condition~\eqref{Assump: Kappa} that is required for the compositionality result. By taking $\varrho_i(s) = s$, $\forall i\in\{1,\ldots,n\}$, the condition~\eqref{Assump: Kappa} and as a result the condition \eqref{compositionality} are always satisfied without any restriction on the number of rooms. Moreover, the compositionality condition~\eqref{compositionality1} is also met since $\beta_i > \eta_i, \forall i\in \{1,\dots,n\}$. Then one can conclude that $\mathds  B(T) = \max_{i} \Big\{0.7659T_i^2 - 30.24T_i + 298.5\Big\}$ is a CBC for the interconnected system $\mathfrak{S}$. Accordingly, $\nu(T) = [ -0.012T_1 + 0.8; \dots;  - 0.012T_{1000} + 0.8]$ is the overall controller for the interconnected system and corresponding parameters satisfying conditions \eqref{sys2}-\eqref{cbceq} are obtained as $\eta=0.13,\beta=4.4,c=0.0139$ and $\kappa(s)=0.99s, \forall s \in \mathbb{R}_{\geq 0}$.

By employing Lemma \ref{reachlem}, we can guarantee that the upper bound on the reachability probability for the element $\vartheta=(q_0,q_1,q_2,9)$ is equal to $0.054$. We now utilize Theorem \ref{th:sumprod} and Corollary \ref{cor:mainprob} to obtain the probability that the specification represented by the accepting language of DFA $\mathcal{A}$ is satisfied by the interconnected system $\mathfrak{S}$ in the time horizon $[0,T_d)$, with $T_d=10$. The lower bound on the probability of the satisfaction of specification represented by  $\mathcal{A}$ when solution processes on the interconnected system start from any initial condition $a \in L^{-1}(p_0)$ is computed as

\begin{equation*}
\mathbb{P}^a_\nu\{L(x_{10}) \models \mathcal{A}\} \geq 0.95.
\end{equation*}

\begin{figure}
	\center
	\includegraphics[scale=0.5]{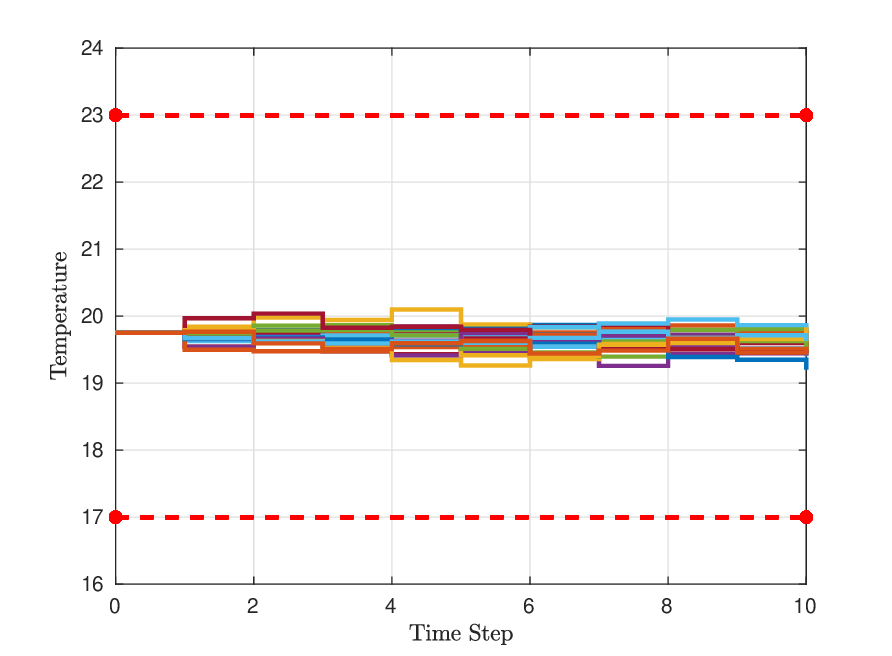}
	\caption{Closed-loop stage trajectories of a representative room with $10$ noise realizations in a network of $1000$ rooms.}
	\label{trajectory1}
\end{figure}

State trajectories of the closed-loop system for a representative room in a network of $1000$ rooms  with $10$ noise realizations are illustrated in Fig.~\ref{trajectory1}. Note that the lower bound on the probability of satisfaction proposed by our approach is rather conservative compared to empirical results that can be obtained by running Monte Carlo simulations for the closed-loop system with our computed controller. The reason is due to the conservative nature of barrier certificates which are chosen to be polynomials of a fixed degree, but at the gain of providing a formal lower bound on the probability of satisfaction rather than just an empirical one. We should mention that the computation of CSBC and its corresponding controller for each subsystem takes almost $10$ seconds with a memory usage of $3.7$ MB on a machine with Microsoft Windows (Intel i7-8665U CPU with a $32$ GB of RAM).

\subsection{Network of Kuramoto Oscillators}

\begin{figure}
	\center
	\includegraphics[scale=0.9]{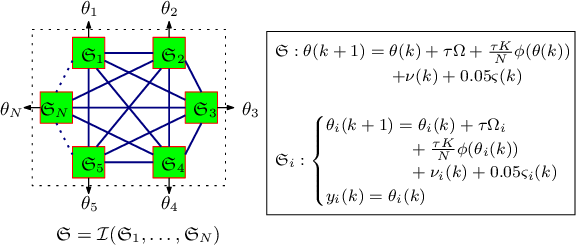}
	\caption{Fully-connected Kuramoto oscillator network $\mathfrak{S}$, and dynamics corresponding to $\mathfrak{S}$ and each subsystem $\mathfrak{S}_i$.}
	\label{fig:kuramoto_network}
\end{figure}

As our second case study, we apply our results to a network of $N=100$ controlled Kuramoto oscillators in a fully-interconnected topology as illustrated in Figure~\ref{fig:kuramoto_network} which can model a large number of problems in different fields, such as biology \cite{cumin_generalising_2007}, smart grids \cite{giraldo_tracking_2014}, neural networks \cite{novikov_oscillatory_2014} and nanotechnology \cite{vodenicarevic_nanotechnology-ready_2017}. Our model is  adapted from \cite{skardal_control_2015} by adding stochasticity as an additive noise and the dynamics of such a model is also presented in Figure~\ref{fig:kuramoto_network}. Here, $\theta=[\theta_1;\ldots;\theta_N]$ is the phase of oscillators with $\theta_i \in [0,2\pi],\ \forall i \in \{1,\ldots,N\}$, $\Omega=[\Omega_1;\ldots;\Omega_N]= [0.01;\ldots;0.01]$ is the natural frequency of oscillators, $K=0.0012$ is the coupling strength, $\tau=0.1$ is the sampling time, $\phi(\theta(k))=[\phi(\theta_1(k));\ldots;\phi(\theta_N(k))]$ such that $\phi(\theta_i(k))=\sum_{\substack{j=1 \\ i \neq j}} sin(\theta_j(k)-\theta_i(k)), \forall i \in \{1,\ldots,N\}$, $\nu(k)=[\nu_1(k);\ldots;\nu_N(k)]$, and $\varsigma(k)=[\varsigma_1(k);\ldots;\varsigma_N(k)]$.
Regions of interest are given by $X^0=[0,\frac{\pi}{15}]^{N}, X^1=[\frac{4\pi}{9},\frac{5\pi}{9}]^N, X^2=[\frac{14\pi}{15},\pi]^N, X^3=[\pi,\frac{16\pi}{15}]^N, X^4=[\frac{13\pi}{9},\frac{14\pi}{9}]^N, X^5=[\frac{29\pi}{15},2\pi]^N$ and $X^6=X\backslash(X^0 \cup X^1 \cup X^2 \cup X^3 \cup X^4 \cup X^5)$. Each region is associated with an element of the atomic proposition given by $\mathcal{AP}=\{p_0,p_1,p_2,p_3,p_4,p_5,p_6\}$ such that the labeling function $L(x_l)=p_l$ for all $x_l \in X^l, l \in \{0,1,\ldots,6\}.$	 

\begin{figure}
	\center
	\includegraphics[width=0.32\linewidth]{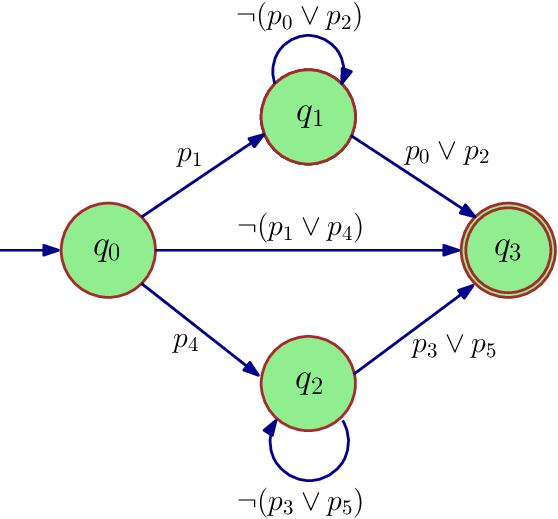}
	\caption{DFA $\mathcal{A}^c$ representing the complement of specification.}
	\label{kuramoto_1}
\end{figure}	

The main goal is to compute a controller such that if the system starts from $X^1$, it must always stay away from $X^0$ and $X^2$, and if it starts from $X^4$, it must always stay away from $X^3$ and $X^5$ within the time horizon $[0,T_d) \subseteq \mathbb{N}$, with $T_d=7$. Such a property can be represented as an LTL specification given by $(p_1 \wedge \square \neg (p_0 \vee p_2)) \vee (p_4 \wedge \square \neg (p_3 \vee p_5))$. It can also be represented by the accepting language of a DFA. Figure~\ref{kuramoto_1} shows the complement DFA $\mathcal{A}^c$. The DFA $\mathcal{A}$ representing the original specification can be readily obtained by switching the non-accepting and accepting states in the figure. We first begin by decomposing the complement of the specification into simple reachability problems. We consider accepting state runs without self-loops with $M=7$. The DFA $\mathcal{A}^c$ has three such accepting state runs and $\mathcal{R}_M=\{(q_0,q_3),(q_0,q_1,q_3),(q_0,q_2,q_3)\}$. For all $p \in \mathcal{AP}$, we have $\mathcal{R}^{p_0}_M=\mathcal{R}^{p_2}_M=\mathcal{R}^{p_3}_M=\mathcal{R}^{p_5}_M=\mathcal{R}^{p_6}_M=\{(q_0,q_3)\}$, $\mathcal{R}^{p_1}_M=\{(q_0,q_1,q_3)\}$, and $\mathcal{R}^{p_4}_M= \{(q_0,q_2,q_3)\}$. Sets $\mathcal{P}^{p}(\textbf{q})$ can be obtained for each of these accepting state runs as $\mathcal{P}^{p_1}(q_0,q_1,q_3)=\{(q_0,q_1,q_3,6)\}$ and $\mathcal{P}^{p_2}(q_0,q_2,q_3)=\{(q_0,q_2,q_3,6) \}$. Note that since $\textbf{q}=(q_0,q_3)$ is a state run of a length 2, it admits a trivial probability as mentioned in Remark \ref{remarkdecomp}, and therefore, it can be neglected. We need to find control barrier certificates and corresponding controllers for the remaining two reachability elements.

To do so, we consider the network of $N$ nonlinear oscillators as an interconnection of $N$ subsystems, \textit{i.e.}, $\mathfrak{S}=\mathcal{I}(\mathfrak{S}_1, \ldots, \mathfrak{S}_N)$  where each subsystem $\mathfrak{S}_i, i \in \{1,\ldots,N\}$, can be described by dynamics as shown in Figure~\ref{fig:kuramoto_network}. To compute control sub-barrier certificates and the corresponding local controllers, we utilize the SOS algorithm in Section \ref{sossubsec} and in particular, we use SOSTOOLS and SDP solver SeDuMi. Since dynamics of $\mathfrak{S}$ are not polynomial and SOS algorithm is only equipped to provide solutions for polynomial dynamics, we make an approximation to our dynamics. More precisely, in the condition~\eqref{sos4}, we take an upper bound on the term $\mathds{B}_i(f_i(\theta_i,\nu_i,w_i,\varsigma_i))$ by replacing $sin(\cdot)$ by either $1$ or $-1$ accordingly.

\begin{figure}
	\centering
	\begin{subfigure}[b]{0.4\textwidth}
		\centering
		\includegraphics[width=1.01\textwidth]{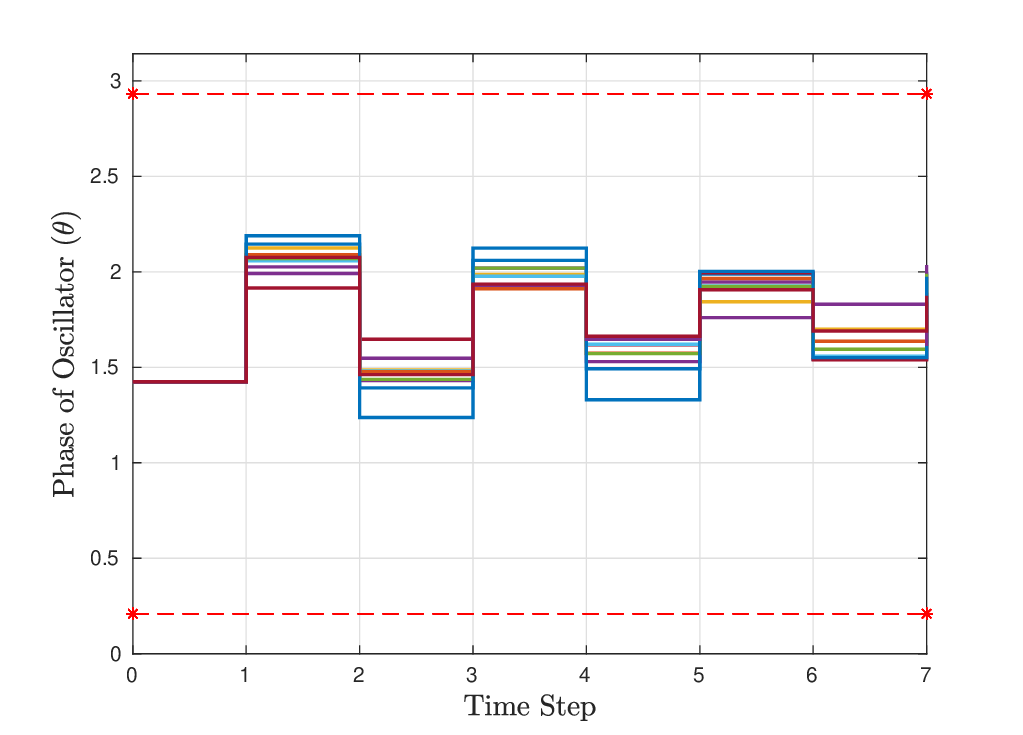}
	\end{subfigure}
	\qquad
	\begin{subfigure}[b]{0.4\textwidth}
		\centering
		\includegraphics[width=\textwidth]{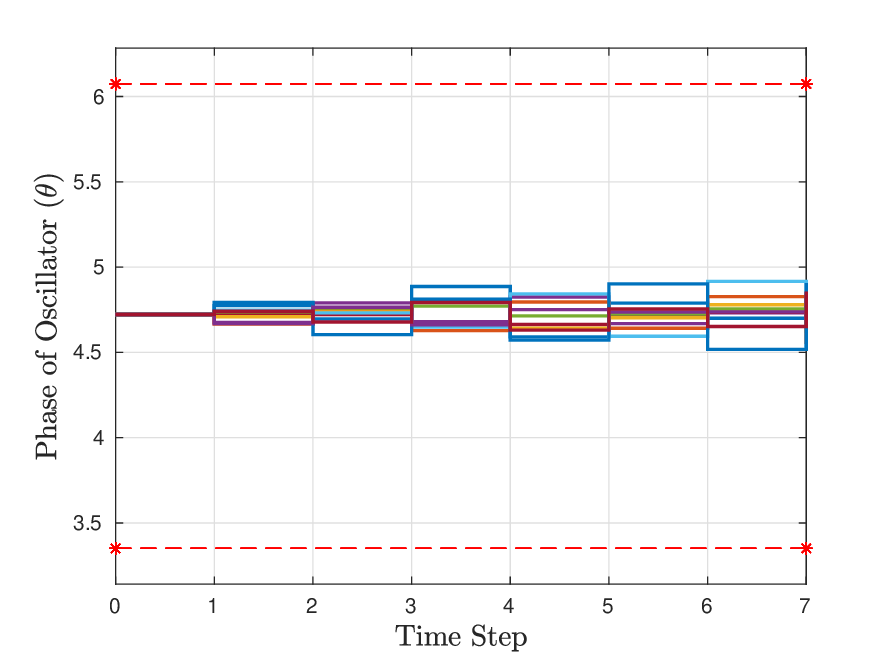}
	\end{subfigure}
	\caption{Closed-loop state trajectories of a representative oscillator in a network of 100 oscillators with 10 noise realizations with an initial state starting from (left) $X_1$, and (right) $X_4$.} 
	
	\label{kuramoto_3}
\end{figure}

The CSBC, local controller and other parameters satisfying conditions \eqref{subsys1}-\eqref{csbceq} for the reachability elements $\vartheta_1$ and $\vartheta_2$ are shown in Table~\ref{tab}. For both elements, one can see that Assumption \ref{Assump: Gamma} is satisfied. Therefore, by utilizing Theorem \ref{Thm: Comp}, we compute CBC and controller for the interconnected system, and also obtain the parameters satisfying \eqref{sys2}-\eqref{cbceq}. Then, by Lemma \ref{reachlem}, we correspondingly obtain  upper bounds for reaching states corresponding to $p_0 \vee p_2$ and $p_3 \vee p_5$ from $p_1$ and $p_4$, respectively. These values are reported in Table~\ref{tab2}. The switching mechanism for controllers is obtained as described in Subsection~\ref{subcontrolpolicy}. Now, by employing Theorem \ref{th:sumprod} and Corollary \ref{cor:mainprob}, we obtain the lower bound on the probability that the solution processes of the interconnected system $\mathfrak{S}$ start from an initial state $a \in X^{1}$ and satisfy the specification represented by the language of DFA $\mathcal{A}$ within the time horizon $T_d=7$ as

\[\mathbb{P}^a_\nu\{L(x_{7}) \models \mathcal{A}\} \geq 0.94.\]
Similarly, for the solution processes of the interconnected system $\mathfrak{S}$ starting from $a \in X^{4}$, we acquire \[\mathbb{P}^a_\nu\{L(x_{7} \models \mathcal{A}\} \geq 0.9.\] Fig.~\ref{kuramoto_3} shows the evolution of solution processes within the time horizon $T_d=7$ when starting from initial regions of $X_1$ and $X_4$.  
The CSBC computation for $\vartheta_{1}$ takes $1$ minute with a memory usage of $30$ MB and for $\vartheta_{2}$, it takes $20$ seconds and $1$ MB memory on a Microsoft Windows machine (Intel i7-8665U CPU with 32 GB of RAM).

\begin{table*}[t]
	\footnotesize
	\caption{CSBC, controller, and parameters obtained for reachability elements $\vartheta$ for all $1 \leq i \leq N$ subsystems.}
	\label{tab}
	\begin{tabular}{@{}cp{10em}cccccccc@{}}
		\toprule 
		$\vartheta$ & $\B_i(\theta_i)$ & $\nu_{i\vartheta}(\theta_i)$ & $\eta_i$ & $\beta_i$ & $c_i$ & $\alpha_i(s)$ & $\kappa_i(s)$ & $\rho_i(s)$ & $\varrho_i(s)$ \\
		\bottomrule 
		\toprule
		$(q_0,q_1,q_3,6)$    & $0.001361\theta_i^8- 
		0.0001877\theta_i^7+ 0.0004904\theta_i^6 - 0.03395\theta_i^5 + 0.00107\theta_i^4 - 0.1927\theta_i^3 + 1.71\theta_i^2 - 3.205\theta_i +  1.827$    & $-0.532\theta^2_i+1.69$   & $0.02$ & $1.2$ &  $0.0083$ & $4.7 \times 10^{-7}s $ & $0.997s $ & $ 4.49 \times 10^{-7}s $ & $s$     \\
		$(q_0,q_2,q_3,6)$  & $0.5396\theta{_i}^2 - 5.086\theta{_i} + 11.86$ &   $-0.21\theta_i^2 + 4.6591$ &  $0.017$ & $1$ & $ 0.0162 $ & $4.5 \times 10^{-8}s$  &   $0.998s$ & $4.49 \times 10^{-8}s$ & $s$   \\   
		\hline     
	\end{tabular}
\end{table*}
\begin{table*}[t]
	\footnotesize
	\centering
	\caption{CBC, controller, and probabilistic guarantees obtained for reachability elements $\vartheta$ for the interconnected system.}
	\label{tab2}
	\begin{tabular}{@{}cp{15em}p{15em}ccccc@{}}
		\toprule 
		$\vartheta$ & $\B(\theta)$ & $\nu_{\vartheta}(\theta)$ & $\eta$ & $\beta$ & $c$ & $\kappa(s)$ & $\varkappa_{\vartheta T_h}$ \\
		\bottomrule 
		\toprule
		$(q_0,q_1,q_3,6)$    & $\max_i \{0.001361\theta_i^8-  0.0001877\theta_i^7+ 0.0004904\theta_i^6 - 0.03395\theta_i^5 + 0.00107\theta_i^4 - 0.1927\theta_i^3 + 1.71\theta_i^2 - 3.205\theta_i +  1.827\}$    & $[-0.532\theta_1^2 + 1.69;\ldots; -0.532\theta_{100}^2 + 1.69 \theta_{100}]$  & $0.02$ & $1.2$ &  $0.0083$  & $0.997s $ & $0.0568$     \\
		$(q_0,q_2,q_3,6)$  & $\max_i\{ 0.5396\theta{_i}^2 - 5.086\theta{_i} + 11.86\}$ &   $[-0.21\theta_1^2 + 4.6591;\ldots;-0.21\theta_{100}^2 + 4.6591\theta_{100}]$ &  $0.017$ & $1$ & $ 0.0162 $ &    $0.998s$ & $0.109$   \\   
		\hline     
	\end{tabular}
\end{table*}

\section{Discussion}

In this paper, we proposed a compositional approach for the verification and synthesis of large-scale interconnected discrete-time stochastic control systems against complex logic specifications that can be described by the accepting language of deterministic finite automata. We first introduced the notion of control sub-barrier certificates which can be utilized to construct control barrier certificates for interconnected systems under some small-gain conditions. Employing those control barrier certificates, we quantified some upper bounds on the probability that interconnected systems reach specified unsafe regions in finite-time horizons. We also provided a systematic approach to decompose a complex specification into a set of reachability tasks using the automaton corresponding to the complement of the specification. We combined probabilities of satisfactions for individual reachability tasks in a sum-product form to obtain a lower bound on the probability of satisfaction of the original specification. In order to compute control sub-barrier certificates for subsystems, we provided two different systematic methods based on sum-of-squares (SOS) optimization problem and counter-example guided inductive synthesis (CEGIS) framework. Finally, applicability of our results were validated via two different large-scale case studies. 

\bibliographystyle{alpha}
\bibliography{ifacconf}

\section{Appendix}

\begin{proof} (\textbf{Theorem \ref{Kushner}})
	According to the condition~\eqref{sys3}, $X_u\subseteq \{x\in X \,\,\big|\,\, \mathds B(x) \ge \beta \}$. Then we have
	\begin{align}
	\mathbb{P}^{a}_{\nu}\Big\{x(k)\in X_u \text{ for } 0\leq k < T_d\,\,\big|\,\, a\Big\}
	\leq\mathbb{P}^{a}_{\nu}\Big\{ \sup_{0\leq k < T_d} \mathds B(x(k))\geq \beta \,\,\big|\,\, a \Big\}. \label{Eq:5}
	\end{align}
	The proposed bounds in~\eqref{eqlemma2} follows directly by applying~\cite[Theorem 3, Chapter III]{1967stochastic} to~\eqref{Eq:5} and employing respectively conditions~\eqref{cbceq} and~\eqref{sys2}.
\end{proof}

\begin{proof} (\textbf{Theorem \ref{Thm: Comp}})
	We first show that conditions~\eqref{sys2} and \eqref{sys3} in Definition \ref{cbc} hold. For any $x\Let[x_{1};\ldots;x_{N}] \in X_0 = \prod_{i=1}^{N} X_{0_i} $ and from \eqref{subsys2}, we have
	\begin{align}\notag
	\mathds B(x) = \max_{i} \Big\{\varrho_i^{-1}(\mathds B_i(x_i))\Big\}\leq \max_{i} \Big\{\varrho_i^{-1}( \eta_i)\Big\} = \eta,
	\end{align} 
	and similarly for any $x\Let[x_{1};\ldots;x_{N}] \in X_u = \prod_{i=1}^{N} X_{u_i} $ and from \eqref{subsys3}, one has
	\begin{align}\notag
	\mathds B(x) = \max_{i} \Big\{\varrho_i^{-1}(\mathds B_i(x_i))\Big\}\geq \max_{i} \Big\{\varrho_i^{-1}(\beta_i)\Big\} = \beta,
	\end{align} 
	satisfying conditions \eqref{sys2} and \eqref{sys3} with $\eta = \max_{i} \Big\{\varrho_i^{-1}( \eta_i)\Big\} $ and $\beta = \max_{i} \Big\{\varrho_i^{-1}(\beta_i)\Big\}$. 
	
	Now we show that the condition~\eqref{cbceq} holds, as well. Let $\kappa(s)= \max_{i,j}\{\varrho_{i}^{-1}\circ\kappa_{ij}\circ\varrho_{j}(s)\}$. It follows from~\eqref{compositionality} that $\kappa<\mathcal{I}_d$. Moreover, $\beta > \eta$ according to~\eqref{compositionality1}. Since $\max_{i}\varrho_{i}^{-1}$ is concave, one can readily acquire the chain of inequalities in \eqref{Equ1b} using Jensen's inequality, and by defining the constant $c$ as
	\begin{align*}
	c:=\max_{i}\varrho_{i}^{-1}(c_{i}).
	\end{align*}	 
	Hence $\mathds B(x)$ is a CBC for the interconnected system $\mathfrak{S}$ which completes the proof.
\end{proof}

	\begin{figure*}[hbt!]
	\rule{\textwidth}{0.1pt}
	\begin{align}\nonumber
	\mathbb{E}\Big[\mathds B(f(x,\nu,\varsigma))\,\big|\,x,{\nu}\Big]&=\mathbb{E}\Big[\max_{i}\Big\{\varrho_{i}^{-1}(\mathds B_{i}(f_i(x_i,\nu_i,w_i,\varsigma_i)))\Big\}\,\big|\,x,{\nu},w\Big]\\\notag
	&\le\max_{i}\Big\{\varrho_{i}^{-1}(\mathbb{E}\Big[\mathds B_{i}(f_i(x_i,\nu_i,w_i,\varsigma_i))\,\big|\,x,{\nu},w\Big])\Big\}\\\notag
	&=\max_{i}\Big\{\varrho_{i}^{-1}(\mathbb{E}\Big[\mathds B_{i}(f_i(x_i,\nu_i,w_i,\varsigma_i))\,\big|\,x_i, \nu_{i},w_i\Big])\Big\}\\\notag
	&\leq\max_{i}\Big\{\varrho_{i}^{-1}(\max\{\kappa_{i}(\mathds B_{i}(x_i)),\rho_{i}(\Vert w_i\Vert^2), c_{i}\})\Big\}\\\notag
	&=\max_{i}\Big\{\varrho_{i}^{-1}(\max\{\kappa_{i}(\mathds B_{i}(x_i)),\rho_{i}(\max_{j, j\neq i}\{\Vert w_{ij}\Vert^2\}), c_{i}\})\Big\}\\\notag
	&=\max_{i}\Big\{\varrho_{i}^{-1}(\max\{\kappa_{i}(\mathds B_{i}(x_i)),\rho_{i}(\max_{j, j\neq i}\{\Vert y_{ji}\Vert^2\}), c_{i}\})\Big\}\\\notag
	&=\max_{i}\Big\{\varrho_{i}^{-1}(\max\{\kappa_{i}(\mathds B_{i}(x_i)), \rho_{i}(\max_{j, j\neq i}\{\Vert h_j(x_j)\Vert^2\}), c_{i}\})\Big\}\\\notag
	&\leq\max_{i}\Big\{\varrho_{i}^{-1}(\max\{\kappa_{i}(\mathds B_{i}(x_i)),\rho_{i}(\max_{j , j\neq i}\{\alpha_{j}^{-1}(\mathds B_{j}(x_j))\}),c_{i}\})\Big\}\\\notag
	&=\max_{i,j}\Big\{\varrho_{i}^{-1}(\max\{\kappa_{ij}(\mathds B_{j}(x_j)),c_{i}\})\Big\}=\max_{i,j}\Big\{\varrho_{i}^{-1}(\max\{\kappa_{ij}\circ \varrho_{j}\circ \varrho_{j}^{-1}(\mathds B_{j}(x_j)),c_{i}\})\Big\}\\\notag
	&\leq\max_{i,j,l}\Big\{\varrho_{i}^{-1}(\max\{\kappa_{ij}\circ \varrho_{j} \circ \varrho_{l}^{-1}(\mathds B_{l}(x_l)),c_{i}\})\Big\}\\\label{Equ1b}
	&=\max_{i,j}\Big\{\varrho_{i}^{-1}(\max\{\kappa_{ij}\circ\varrho_{j}\circ\mathds B(x),c_{i}\})\Big\}=\max\Big\{\kappa(\mathds B(x)),c\Big\}.
	\end{align}
	\rule{\textwidth}{0.1pt}
\end{figure*}

\begin{proof} (\textbf{Theorem \ref{th:sumprod}})
	Consider a set of accepting state runs $\mathcal{R}^{p}_M$ of the length of at most $M+1$ for all $p \in \mathcal{AP}$ and set $\mathcal{P}^p(\textbf{q})$ as the set of state runs of a length $3$, augmented with the horizon $T_h$. For $\vartheta=(q,q',q'',T_h) \in \mathcal{P}^p(\textbf{q})$, we can establish from Lemma \ref{reachlem} that the upper bound on the probability that a solution process of dt-SCS $\mathfrak{S}$ starts from $X_0=L^{-1}(\sigma(q,q'))$ and reaches $X_u=L^{-1}(\sigma(q',q''))$ within the time horizon $[0,T_h) \subseteq \mathbb{N}$ under the influence of the control input $\nu_{\vartheta}$ is given by $\varkappa_{\vartheta T_h}$.  Then the probability that the solution process reaches the accepting state by following the finite word corresponding to $\textbf{q}$ is the product of all probability bounds corresponding to elements $\vartheta=(q,q',q'',T_h) \in \mathcal{P}^p(\textbf{q})$ and is given by
	
	\begin{equation*} 
	\mathbb{P}^a_\nu(L(x_M) \models \mathcal{A}^c ) \leq  \prod_{\vartheta \in \mathcal{P}^p(\textbf{q})}\hspace{-1em}\{\varkappa_{\vartheta T_h} \,\big|\, \vartheta=(q,q',q'',T_h) \in \mathcal{P}^{p}(\textbf{q})\}.
	\end{equation*}  
	Using the horizon $T_h$, we obtain the upper bound on probabilities for accepting state runs $\mathcal{R}^p_M$ with a length of at most $M+1$  by considering all possible self-loop combinations. Given the initial condition $a \in L^{-1}(p)$, the final upper bound for a solution process of $\mathfrak{S}$ to violate the required specification is essentially the summation of probabilities of all possible accepting state runs from the initial state to the final state of $\mathcal{A}^c$, and is given by
	
	\begin{align*}
	&\mathbb{P}^a_\nu\{L(x_M) \models \mathcal{A}^c\}  \leq   \sum_{q \in \mathcal{R}^p_M}  \prod_{\vartheta \in \mathcal{P}^p(\textbf{q})}\{\varkappa_{\vartheta T_h} \,\big|\, \vartheta  =  (q,q',q'',T_h) \in  \mathcal{P}(\textbf{q})\}, 
	\end{align*}	
	which completes the proof.
\end{proof}

\begin{proof} (\textbf{Lemma \ref{sos}})
	Since $\mathbb B_i(x_i)$ and $\lambda_i(x_i)$ in~\eqref{sos1} are sum-of-squares, we have $0\leq \mathbb B_i(x_i)- \lambda_{i}^T(x_i) g_{i}(x_i)-\alpha_i(h_i^T(x_i)h_i(x_i))$. Since $\norm{h_i(x_i)}^2 \leq h_i^T(x_i)h_i(x_i)$, we have  $0\leq \mathbb B_i(x_i)- \lambda_{i}^T(x_i) g_{i}(x_i)-\alpha_i(\Vert h_i(x_i)\Vert^2)$. Since the term $\lambda_{i}^T(x_i) g_{i}(x_i)$ is non-negative over $X$, the condition~\eqref{sos1} implies the condition~\eqref{subsys1}.
	Similarly, we can show that \eqref{sos2} and \eqref{sos3} imply conditions \eqref{subsys2} and~\eqref{subsys3}, respectively. Now we proceed with showing that the condition~\eqref{sos4} implies~\eqref{csbceq}, as well. By selecting external inputs $\nu_{ji}=\lambda_{\nu_{ji}}(x_i)$ and since the terms $\hat\lambda^{T}_i(x_i,\nu_i,w_i)g_i(x_i), \lambda_{\nu_i}^{T}(x_i,\nu_i,w_i)g_{\nu_i}(\nu_i), \lambda_{w_i}^{T}(x_i,\nu_i,w_i)g_{w_i}(w_i)$ are non-negative, we have $\mathbb{E}\Big[\mathds B_i(f_i(x_i,\nu_i,w_i,\varsigma_i)) \,\big|\, x_i,\nu_i,w_i\Big] \leq$ $\bar\kappa_i(\mathds B_i(x_i))+ \bar\rho_{i}(\frac{w_i^Tw_i}{p_i})+\bar c_i$ implying that $\mathbb{E}\Big[\mathds B_i(f_i(x_i,\nu_i,w_i,\varsigma_i)) \,\big|\, x_i,\nu_i,w_i\Big] \leq$ $\bar\kappa_i(\mathds B_i(x_i))+ \bar\rho_{i}(\norm{w_i}^2)+\bar c_i$, since $w_i^Tw_i \leq p_i \norm{w_i}^2$. By employing a similar argument as the one in~\cite[Theorem 1]{Abdallah2017compositional}, the additive form of the right-hand side of the above inequality can be converted into a max form as
	\begin{align}\notag
	\mathbb{E}&\Big[\mathds B_i(f_i(x_i,\nu_i,w_i,\varsigma_i)) \mid x_i,\nu_i,w_i\Big]\leq\max\Big\{{\kappa}_i(\mathds B_i(x_i)),\rho_i(\Vert w_i\Vert^2), c_i\Big\},
	\end{align}
	where, 
	\begin{align}\notag
	&{\kappa}_i =\mathcal{I}_d -(\mathcal{I}_d - {\pi}_i )\circ (\mathcal{I}_d - \bar\kappa_i),\\\notag
	&\rho_i=(\mathcal{I}_d + \bar {\delta}_i)\circ(\mathcal{I}_d - \bar\kappa_i )^{-1}\circ {\pi}_i ^{-1}\circ\bar\pi_i\circ\bar \rho_i,\\\notag
	&c_i=(\mathcal{I}_d + \bar {\delta}_i^{-1})\circ(\mathcal{I}_d - \bar\kappa_i)^{-1}\circ {\pi}_i  ^{-1}\circ\bar\pi_i\circ(\bar \pi_i - \mathcal{I}_d)^{-1} ( \bar c_i),
	\end{align}
	with $\bar {\delta}_i, {\pi}_i ,\bar\pi_i$ being some arbitrarily chosen $\mathcal{K}_\infty$ functions so that $\mathcal{I}_d - {\pi}_i  \in \mathcal{K}_\infty$, $\bar\pi_i - \mathcal{I}_d \in \mathcal{K}_\infty$. Hence this implies that the function $\mathds B_i(x_i)$ is a CSBC and the proof is completed.
\end{proof}

\end{document}